\documentclass[aps,nofootinbib,superscriptaddress,twocolumn,floatfix,longbibliography]{revtex4-1}
\usepackage{mathrsfs}
\usepackage{graphicx}
\usepackage{amsmath,amssymb}
\usepackage{hyperref}
\usepackage{braket}
\usepackage{subfigure}
\usepackage{subfloat}
\usepackage{float}
\usepackage[usenames]{color}

\hypersetup{
    colorlinks=true,
    linkcolor=red,
    citecolor=blue,
}

\allowdisplaybreaks

\def\be{\begin{equation}}
\def\ee{\end{equation}}
\def\ba{\begin{eqnarray}}
\def\ea{\end{eqnarray}}

\newcommand{\lcdm}{$\Lambda$CDM}

\begin{document}

\title{Imprints of cosmological tensions in reconstructed gravity}

\author{Levon Pogosian} \affiliation{Department of Physics, Simon Fraser University, Burnaby, BC, V5A 1S6, Canada}
\affiliation{Institute of Cosmology and Gravitation, University of Portsmouth, Portsmouth, PO1 3FX, UK}
\affiliation{National Astronomy Observatories, Chinese Academy of Science, Beijing, 100101, P.R.China}
\email[]{levon@sfu.ca}

\author{Marco Raveri} 
\affiliation{Center for Particle Cosmology, Department of Physics and Astronomy, University of Pennsylvania, Philadelphia, PA 19104, USA}
\email[]{mraveri@sas.upenn.edu}

\author{Kazuya Koyama}
\affiliation{Institute of Cosmology and Gravitation, University of Portsmouth, Portsmouth, PO1 3FX, UK}

\author{Matteo Martinelli}
\affiliation{INAF - Osservatorio Astronomico di Roma, via Frascati 33, 00040 Monteporzio Catone (Roma), Italy}
\affiliation{Instituto de F\'isica Teorica UAM-CSIC, Campus de Cantoblanco, E-28049 Madrid, Spain}

\author{Alessandra Silvestri} 
\affiliation{Institute Lorentz, Leiden University, PO Box 9506, Leiden 2300 RA, The Netherlands}

\author{Gong-Bo Zhao}
\affiliation{National Astronomy Observatories, Chinese Academy of Science, Beijing, 100101, P.R.China}
\affiliation{University of Chinese Academy of Sciences, Beijing, 100049, P.R.China}

\author{Jian Li}
\affiliation{Department of Physics, Simon Fraser University, Burnaby, BC, V5A 1S6, Canada}

\author{Simone Peirone}\affiliation{Institute Lorentz, Leiden University, PO Box 9506, Leiden 2300 RA, The Netherlands}

\author{Alex Zucca} \affiliation{Department of Physics, Simon Fraser University, Burnaby, BC, V5A 1S6, Canada}

\begin{abstract} 
There has been a significant interest in modifications of the standard $\Lambda$ Cold Dark Matter ($\Lambda$CDM) cosmological model prompted by tensions between certain datasets, most notably the Hubble tension. The late-time modifications of the $\Lambda$CDM model can be parametrized by three time-dependent functions describing the expansion history of the Universe and gravitational effects on light and matter in the Large Scale Structure. We perform the first joint Bayesian reconstruction of these three functions from a combination of recent cosmological observations, utilizing a theory-informed prior built on the general Horndeski class of scalar-tensor theories. This reconstruction is interpreted in light of the well-known $H_0$, the $S_8$ and the $A_L$ tensions. We identify the phenomenological features that alternative theories would need to have in order to ease some of the tensions, and deduce important constraints on broad classes of modified gravity models. Among other things, our findings suggest that late-time dynamical dark energy and modifications of gravity are not likely to offer a solution to the Hubble tension, or simultaneously solve the $A_L$  and $S_8$ tensions. 
\end{abstract}
\maketitle

Despite the success of the $\Lambda$ Cold Dark Matter ($\Lambda$CDM) model in fitting a multitude of cosmological data, there are good reasons to keep an open mind about its possible extensions. Chief among them is the lack of satisfactory understanding of its two key ingredients: $\Lambda$ and CDM. Another reason is the improving depth and resolution of cosmological surveys across a wide range of wavelengths, opening qualitatively new ways of testing the model and some of its fundamental principles, such as the validity of General Relativity (GR) over cosmological distances~\cite{Silvestri:2009hh,Joyce:2014kja,Koyama:2015vza}. Finally, with the increased constraining ability of different types of measurements, a few notable tensions have recently emerged between some of the datasets when interpreted within $\Lambda$CDM.

Most actively discussed is the ``Hubble tension'', referring to the 5$\sigma$ disagreement between the value of the Hubble constant predicted by the $\Lambda$CDM model fit to cosmic microwave background (CMB) measurements~\cite{Planck:2018vyg} and the most recent direct determination from Cepheid calibrated type IA Supernova (SN)~\cite{Riess:2021jrx}. The $H_0$ obtained using alternative methods of calibrating SN have larger uncertainties at this time but promise to become much more precise in the future. They also yield higher values of $H_0$ \cite{Abdalla:2022yfr}, although some are in a better agreement with CMB \cite{Freedman:2020dne,Freedman:2021ahq}.  Another well-known tension is the disagreement in the galaxy clustering amplitude, quantified by the parameter $S_8$, predicted by the best fit to CMB and that measured by galaxy weak  lensing surveys, such as the Dark Energy Survey (DES)~\cite{DES:2021wwk}, the Kilo-Degree Survey (KiDS)~\cite{Asgari:2020wuj} and the Subaru Hyper Suprime-Cam (HSC) survey \cite{HSC:2018mrq}. In addition, the CMB temperature anisotropy measured by Planck appears to be more affected by weak gravitational lensing than expected in $\Lambda$CDM~\cite{Planck:2018vyg}. Various extensions of $\Lambda$CDM have been proposed with the aim of relieving some of these tensions, including modifications of GR~\cite{Abdalla:2022yfr}. The possible role of systematics in these tensions has also been studied~\cite{Efstathiou:2017rgv, Efstathiou:2020wxn}. 

GR is one of the most successful physical theories, having passed many tests in Earth-based laboratories and the Solar system~\cite{Will:2014kxa} and, more recently, having been validated by observations of gravitational waves from binary black holes and neutron stars~\cite{Abbott:2016blz,TheLIGOScientific:2017qsa}, and the imaging of the black hole in M87~\cite{EventHorizonTelescope:2019dse}. None of these tests, however, probe GR on cosmological scales, where gravity, rather than being sourced by massive objects, is characterized by the Hubble expansion of the universe. The discovery of the acceleration of cosmic expansion~\cite{Riess:1998cb,Perlmutter:1998np}, which, within GR, implies the existence of mysterious dark energy (DE), along with the older puzzle concerning the vacuum energy and whether it gravitates ({\it i.e.} the old cosmological constant problem~\cite{Burgess:2013ara}), prompted significant interests in possible alternatives of GR, commonly referred to as modified gravity (MG). While no preferred alternative has emerged so far, significant progress has been made over the past decade and a half in identifying the key requirements of a successful theory. Many of the earlier MG models were shown to be specific realizations from a general class of scalar-tensor theories with second order equations of motion, discovered by Horndeski in 1974~\cite{Horndeski:1974wa}. A broad understanding was also achieved of the possible ways in which MG effects can be screened in order to comply with the many stringent tests of GR~\cite{Vainshtein:1972sx,Damour:1994zq,Khoury:2003aq,Hinterbichler:2010es,Joyce:2014kja}. The resulting theoretical landscape is rich and complex, but the emergence of phenomenological and unifying frameworks~\cite{Amendola:2007rr,Bertschinger:2008zb,Pogosian:2010tj,Gubitosi:2012hu,Bloomfield:2012ff,Gleyzes:2014rba,Bellini:2014fua}, and their numerical implementations~\cite{Zhao:2008bn,Hojjati:2011ix,Hu:2013twa,Zumalacarregui:2016pph}, helped to identify the key signatures and sets of promising cosmological probes.

What can cosmology tell us about gravity? The observable universe is homogeneous and isotropic on large scales, well-described by the flat Friedmann-Lemaitre-Robertson-Walker (FLRW) metric with line element $ds^2 = -dt^2 + a(t)^2d{\bf x}^2 $, where $a(t)$ is the scale factor describing the background expansion. The evolution of the latter is determined by the Friedmann equation, which can, in general, be written as
\be\label{Friedmann}
\left({da \over dt}\right)^2 = a^2 H_0^2 \left[ {\Omega_r \over a^4}+ {\Omega_m \over a^3}+ \Omega_X(a) \right] \,,
\ee
where $H_0$ is the Hubble constant, $\Omega_r$ and $\Omega_m$ are the current fractional energy density in relativistic and non-relativistic particle species, and $\Omega_X(a)\equiv \rho_X(a)/\rho_{\rm today}^{\rm critical}$ represents the effective DE density that has a current fraction $\Omega_{\rm DE}$, {\it i.e.}, $\Omega_X(a)=\Omega_{\rm DE}$ at $a=1$, so that, at present, we have $\Omega_r+\Omega_m+\Omega_{\rm DE}=1$.  In $\Lambda$CDM, DE corresponds to the cosmological constant, $\Lambda$, and $\Omega_X(a)={\rm const}$; more generally, $\Omega_X(a)$ describes the collective contribution of any terms other than the radiation and matter densities, including modifications to gravity that would imply a modified Friedmann equation and the possibility of a non-zero curvature term, $\Omega_k/a^2$. Checking if $\Omega_X(a)={\rm const}$ throughout the history of the universe is a key test of the flat $\Lambda$CDM model. Most studies of DE in the literature do so focusing on the equation of state of DE, $w(a)=(1/3)d\ln{\Omega_X(a)}/d\ln{a}-1$, looking for departures from $w_\Lambda=-1$. In MG theories, however, the effective DE density can pass through zero, making $w$ singular.

On smaller scales, inhomogeneities become important. In the Newtonian gauge, focusing on scalar components, the perturbed FRW line element reads
\begin{equation}
ds^2 = -(1+2\psi)dt^2 + a^2(1-2\phi)d{\bf x}^2 \ ,
\end{equation}
where $\psi=\psi(t,{\bf x})$ and $\phi=\phi(t,{\bf x})$ correspond to the Newtonian potential and spatial curvature inhomogeneity, respectively. A theory of gravity, such as GR, provides a set of equations that relate these metric perturbations to the inhomogeneities in matter. At linear order, they can be written in Fourier space as (neglecting the anisotropic stress contribution from radiation)
\ba
k^2 \psi &=& -4\pi G \mu(a,k)a^2 \rho \Delta \ ,\\
\label{poisson_mg}
k^2(\phi +\psi)/2 &=& -4 \pi G\Sigma(a,k) a^2  \rho \Delta   \ ,
\label{shear_mg}
\ea
where $G$ is the Newton's constant, $k$ is the Fourier wavenumber, $\rho$ is the background density of matter and $\Delta$ is the comoving matter density contrast. The phenomenological functions $\mu$ and $\Sigma$ are defined by the relations above, and are both equal to unity in $\Lambda$CDM. The Weyl potential, $(\phi +\psi)/2$, determines the trajectories of light, probed by weak gravitational lensing (WL) of galaxies or CMB, while $\psi$ is the potential felt by non-relativistic matter, determining the peculiar velocity of galaxies observed via redshift space distortions (RSD). Thus, combining WL and RSD probes, along with other cosmological data, provides a way of measuring $\mu$ and $\Sigma$~\cite{Amendola:2007rr,Pogosian:2010tj,Song:2010fg}. A related commonly used phenomenological function is the gravitational slip,
\ba
\gamma \equiv \frac{\phi}{\psi} =  \frac{2\Sigma}{\mu} -1 \ ,
\label{eq:slip}
\ea
which is equal to unity in GR. The slip is a ``smoking gun'' of MG, as any evidence of $\mu \ne \Sigma$, or $\gamma \ne 1$, would signal a breakdown of the equivalence principle -- a key prediction of GR. The slip is also intimately related to the speed of gravitational waves, $c_T$~\cite{Saltas:2014dha}. In addition, any departure of $\mu$ or $\Sigma$ from unity would be a signature of new interactions or particle species. Furthermore, broad subsets of the Horndeski class of theories can be ruled out depending on the measured values of $\mu$ and $\Sigma$~\cite{Pogosian:2016pwr}.

In Horndeski theories, $\mu$ and $\Sigma$ are ratios of second order polynomials in $k$ \cite{Silvestri:2013ne}, with the $k$-dependence set by the Compton wavelength of the scalar field. For scalar-tensor theories to be viable, while still having cosmological signatures, they must include a screening mechanism that restores GR in the Solar System. There are two broad types of screening mechanisms: Vainshtein and Chameleon~\cite{Vainshtein:1972sx,Damour:1994zq,Khoury:2003aq,Hinterbichler:2010es,Joyce:2014kja}. The Compton length tends to be either comparable to the Hubble scale, in the former case, or below 1 Mpc in the latter. Since, either way, the scale-dependence is outside the range probed by large scale structure surveys within the linear perturbation theory, we do not consider the $k$-dependence of $\mu$ and $\Sigma$, focusing solely on their evolution with redshift. This also helps to reduce the computational costs.



\begin{figure}[tbph!]
\includegraphics[width=\columnwidth]{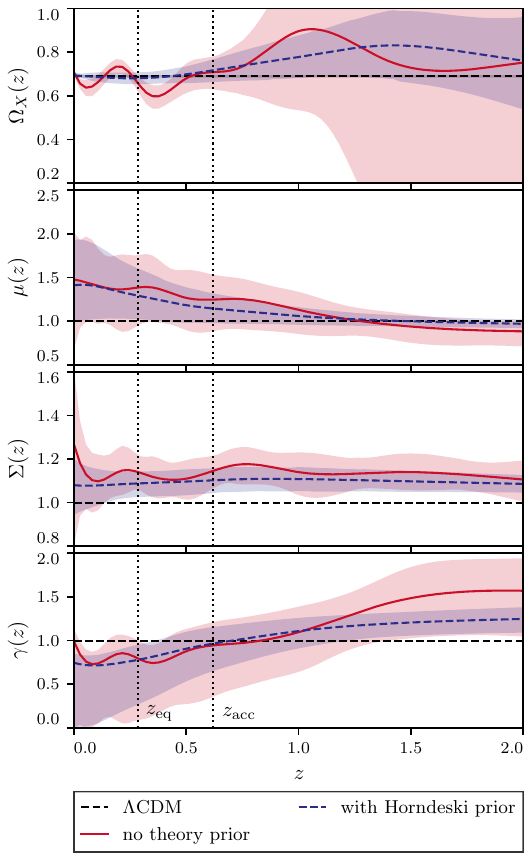}
\caption{\label{fig:all_functions} Reconstructed evolution of $\Omega_X(z)$, $\mu(z)$ and $\Sigma(z)$, along with the derived reconstruction of $\gamma(z)$, from the Baseline+LSS data, with and without the Horndeski prior. The bands show the $68\%$ confidence level regions derived from the marginalized posterior distributions of the individual nodes. The two vertical lines show the redshifts of equality between the matter and DE densities, $z_{\rm eq}$, and the beginning of cosmic acceleration, $z_{\rm acc}$, in the best fit $\Lambda$CDM model.}
\end{figure}

Fig.~\ref{fig:all_functions} shows the joint reconstruction of $\Omega_X(z)$, $\mu(z)$ and $\Sigma(z)$ from a combination of CMB, Baryon Acoustic Oscillations (BAO), SN, WL and RSD data, along with the derived reconstruction of $\gamma(z)$. The reconstruction is performed with and without a theoretical prior, derived previously from simulations of Horndeski theories \cite{Espejo:2018hxa}.  One can clearly see the important role played by the theory prior in preventing over-fitting the data. This is particularly true for $\Omega_X(z)$,  which exhibits oscillations at $0<z<0.6$ driven by the scatter in the BAO and SN data, which are then completely suppressed by the prior. Overall, the reconstructed $\Omega_X(z)$ is consistent with the $\Lambda$CDM prediction, especially with the prior. 

The signal-to-noise ratio of the detection of deviation from $\Lambda$CDM (see Methods for details) is 2.9 (1.3), 1.7 (1.6), 2.4 (2.3) and 1.9 (1.8) for $\Omega_{X}$, $\mu$, $\Sigma$ and $\gamma$, respectively, without the prior (with the prior). The total $\chi^2$ value is improved by 16.5 (3.9) compared with $\Lambda$CDM without the prior (with the prior). The significance of the detection generally drops after including the Horndeski prior, most notably for $\Omega_X$. The most persistent deviation is in $\Sigma$ driven by the CMB lensing anomaly, as discussed in the following section.

The reconstructed evolution of $\mu$ in Fig.~\ref{fig:all_functions} shows a clear preference for $\mu>1$ at low $z$, and a weak trend for $\mu<1$ at higher $z$. The $\mu>1$ trend is caused by the positive correlation between $\mu$ and $\Sigma$, while the reasons for $\Sigma>1$ will be discussed in the next subsection in the context of the CMB lensing tension. Since $\mu$ directly affects the growth of gravitational potential, the $\mu<1$ trend at higher $z$ is there to compensate for $\mu>1$ at lower $z$ in order to keep the clustering amplitude $S_8$ consistent with the data. 


\begin{figure}[tbph!]
\includegraphics[width=\columnwidth]{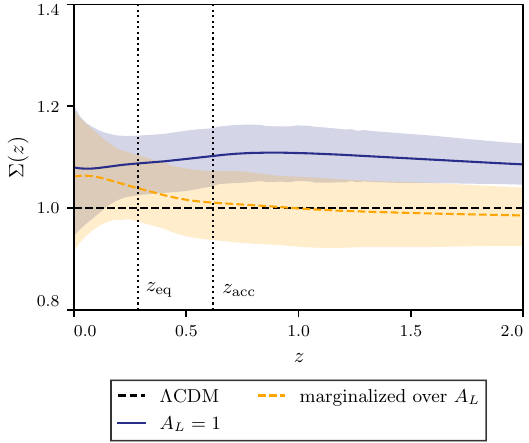}
\caption{\label{fig:sigma} Reconstruction of $\Sigma(z)$ from Baseline+LSS, with the Horndeski prior, with and without varying the TT lensing amplitude parameter $A_{\rm L}$.}
\end{figure}

The reconstructed shape of $\Sigma(z)$ shows interesting departures from $\Lambda$CDM, generally preferring $\Sigma>1$. The increase in $\Sigma$ at lower redshifts helps to better fit the large scale power deficit in the CMB temperature anisotropy spectra (TT), as it reduces the Integrated Sachs Wolfe (ISW) effect by slowing the decay of the gravitational potentials caused by cosmic acceleration. At higher redshifts,  $\Sigma>1$ is caused by the weak lensing anomaly in the Planck TT~\cite{Planck:2018vyg}, with the larger $\Sigma(z)$ helping to boost the Weyl potential responsible for smoothing of the acoustic peaks in TT. However, as one can see from Table~\ref{tab:parametersBaseline} of Supplemental Information (SI), this worsens the fit to the CMB lensing part of Planck (increased $\chi_{\rm CMBL}^2$), suggesting that the anomaly is caused by something other than a deficit of weak lensing. Performing our analysis while allowing the TT lensing amplitude $A_{\rm L}$~\cite{Calabrese:2008rt} to be a free parameter brings the high $z$ values of $\Sigma(z)$ back to unity, as one can see from Fig.~\ref{fig:sigma}. Additional discussion of the lensing anomaly and the other tensions, and their impact on the reconstruction is presented in the SI Section \ref{sec:tensions}.

\begin{figure}[tbph!]
\includegraphics[width=\columnwidth]{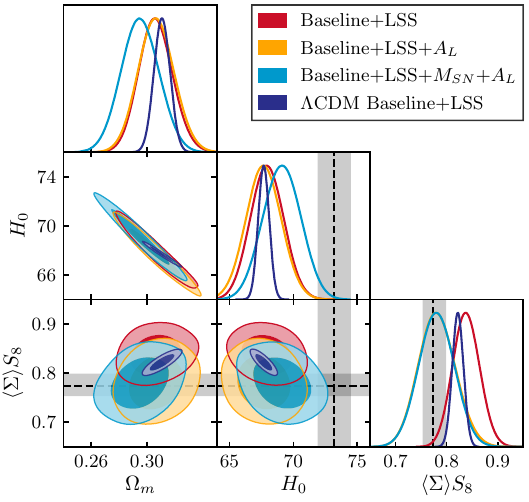}
\caption{\label{fig:tension} The $68\%$ and $95\%$ confidence level contours for $\Omega_M$, $\langle \Sigma \rangle S_8$ and $H_0$ derived from the combination of all data with the Horndeski prior, with and without the SH0ES prior on $M_{\rm SN}$ and the added TT lensing anomaly parameter $A_{\rm L}$. The blue contours show the $\Lambda$CDM best fit, while the grey bands show the constraint on $H_0$ obtained by the SH0ES collaboration and $\langle \Sigma \rangle S_8$ obtained by DES.}
\end{figure}

The analysis above shows that MG can, in principle, help to solve the TT lensing anomaly if $\Sigma >1$. However, this has an important implication for the $S_8$ tension as shown in Fig.~\ref{fig:tension}. The matter clustering amplitude quantified by the $S_8$ parameter is related to the amplitude of the Weyl potential, which in our analysis is directly impacted by $\Sigma$. Thus, the parameter combination that is best constrained by WL surveys, such as DES, is $\langle \Sigma \rangle S_8$, where $\langle \Sigma \rangle$ is the value averaged over the redshift range probed by the given survey. Thus, while MG can lower $S_8$ by allowing for $\Sigma>1$ to fit the TT lensing anomaly, $\langle \Sigma \rangle S_8$ would still remain in tension with the DES value. However, if one ``solves'' the CMB lensing anomaly by allowing for $A_{\rm L}$ to be free, $\Sigma$ returns to its GR value as shown in Fig.~\ref{fig:sigma}, and the lower $S_8$ value restores the agreement with DES for $\langle \Sigma \rangle S_8$. This implies that MG cannot solve the TT lensing anomaly and the $S_8$ tension simultaneously.

\begin{figure}[tbph!]
\includegraphics[width=\columnwidth]{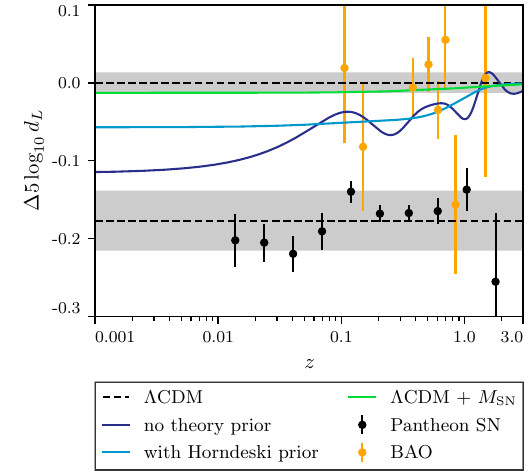}
\caption{\label{fig:shoes} The luminosity distances predicted by the Baseline best fit $\Lambda$CDM (shown as a grey band around $0$) compared to the Pantheon SN data calibrated using the SH0ES measurement of $M_{\rm SN}$. The lower grey band shows the errors in the measurements of $M_{\rm SN}$. Also shown are the BAO data points with the standard ruler calibrated on the Baseline $\Lambda$CDM model. The clear separation between the two grey bands illustrates the Hubble tension, while the gap between the two clusters of data demonstrates the difficulty of solving it with dynamical DE without violating the agreement with either BAO or SN.}
\end{figure}

What about the $H_0$ tension? Fig.~\ref{fig:shoes} demonstrates the general difficulty in solving the Hubble tension by allowing for an evolving DE at low redshifts. In particular, one can see the clear separation between the BAO data cluster, aligned along the upper grey band, corresponding to the luminosity distances calibrated on the Planck best-fit $\Lambda$CDM model, {\it i.e.}, using the sound horizon determined by Planck CMB measurements, and the SN data cluster, aligned along the lower grey band, corresponding to the Cepheid based calibration by SH0ES. Adding the SH0ES prior on $M_{\rm SN}$ to the data brings the reconstructed expansion history closer to the lower band, but it is still far from a good agreement. The posteriors for $H_0$ are shown in Fig.~\ref{fig:tension}. We find $H_0=69. 7 \pm 1.4$ km/s/Mpc from the fit to Baseline+LSS+$M_{\rm SN}$ with the Horndeski prior and $A_L$, which is within $2\sigma$ of the SH0ES value of $H_0=73.2 \pm 1.3$ km/s/Mpc. However, as one can see from Table~\ref{tab:parametersAll} in SI, the value of $\chi_{M}^2$ remains high and the ``easing'' of the tension comes at the cost of a significant increase in the uncertainty of $H_0$.


One can make several useful deductions about MG theories by comparing our reconstructions to the expressions for $\mu$, $\Sigma$ and $\gamma$ derived in Horndeski theories under the quasi-static approximation. It is helpful to consider the expression for $\gamma(z)$ in the small and large $k$ limits~\cite{Gleyzes:2015rua,Pogosian:2016pwr},
\ba
\gamma  \rightarrow 
\begin{cases}
1/c_T^{2}, & k \rightarrow 0 \\
(1+\beta_B \beta_\xi)/(c_T^2 + \beta_\xi^2), & k \rightarrow \infty
\end{cases} \ ,
\ea
where $c_T$ is the speed of gravitational waves, while $\beta_B$ and $\beta_\xi$ represent two different ways of coupling the metric and the scalar field, manifested as a ``fifth'' force felt by matter particles. These limiting expressions show why the gravitational slip is considered a smoking gun of MG. In the context of Horndeski theories, $\gamma \ne 1$ can be due to a modified $c_T$~\cite{Saltas:2014dha}, a fifth force, or both. Since the multi-messenger observations of gravitational waves from binary neutron stars ~\cite{TheLIGOScientific:2017qsa,Monitor:2017mdv} found $c_T=1$ at $z \sim 0$, one would conclude that the observed $\gamma \ne 1$ implies evidence for a fifth force at low redshifts.

The reconstructed $\gamma(z)$ in the bottom panel of Fig.~\ref{fig:all_functions} evolves from $\gamma>1$ at higher $z$ to $\gamma<1$ at lower $z$, with the transition happening around the epoch of onset of cosmic acceleration. While the significance of the $\gamma \ne 1$ detection is less than $2 \sigma$, it is still interesting to consider its implications for Horndeski theories. It would rule out models with $\gamma=1$, such as the Cubic Galileon (CC)~\cite{Deffayet:2009wt}, Kinetic Gravity Braiding (KGB)~\cite{Deffayet:2010qz} or the ``no-slip gravity'' (NSG)~\cite{Linder:2018jil}. In addition, finding $\gamma > 1$ at any redshift would rule out the generalized Brans-Dicke models (GBD), {\it i.e.} all theories with a canonical form of the scalar field kinetic energy term. In GBD, one has $c_T=1$ and $\beta_{B} = - \beta_{\xi}$, and hence one should have $\gamma \le 1$ on all scales. 

In addition, we find no violation of the $(\Sigma-1)(\mu-1) \ge 0$ condition expected in Horndeski theories~\cite{Pogosian:2016pwr}, implying that models with non-canonical kinetic terms are permitted.


Until the nature of DE and CDM are properly understood, we should use every opportunity for testing the basic assumptions of the model, including the validity of GR on cosmological scales. Our methodology is readily applicable to the data from upcoming surveys which will make it possible to reconstruct gravity on cosmological scales at unprecedented precision, providing a stringent test of the standard model of cosmology. We have identified the phenomenological features that alternative gravity and DE theories would need to have in order to ease some of the tensions present within the \lcdm \ model. Overall, our results suggest that, while theories of late time modifications of gravity can help ease some of the tensions, they are unlikely to eliminate all of them simultaneously. 
We also observed hints of departures of the gravitational slip from its GR value of $\gamma=1$ which, if confirmed at a higher statistical significance by future observations, would constitute a smoking gun of modified gravity, while ruling out several popular MG theories.

\section{Methods}
\label{Sec:Methods}
 
\subsection{Reconstruction and the role of the theory prior}

In the reconstruction of $\Omega_X$, $\mu$ and $\Sigma$, each of the three functions is represented by its values at 11 points in $a$: 10 values (nodes) uniformly spaced in $a\in [1,0.25]$ (corresponding to $z\in [0,3]$) and an additional node at $a=0.2$ ($z=4$), with a cubic spline connecting the nodes and making the functions approach their GR values at $a \rightarrow 0$. The cubic spline introduces a correlation of the $11$ redshift nodes, making the statistical significance of the reconstruction dependent on an implicit smoothing scale determined by our arbitrary choice of the number of nodes. This ambiguity is eliminated after the data is supplemented with a theoretical correlation prior, as long as the smoothness scale imposed by the prior is larger than that of the cubic spline. This is indeed the case in our reconstruction, where we used the prior was derived in~\cite{Espejo:2018hxa} by generating large ensembles of solutions in Horndeski theories within the Effective Field Theory (EFT) framework and projecting them onto $\Omega_X(a)$, $\mu(a)$ and $\Sigma(a)$. The most important feature of the Horndeski prior is a strong positive correlation between $\mu$ and $\Sigma$, with preference for $(\Sigma-1)(\mu-1) \ge 0$, which was anticipated in~\cite{Pogosian:2016pwr} based on analytical considerations and later confirmed by a numerical sampling of Horndeski solutions~\cite{Peirone:2017ywi,Espejo:2018hxa}.

These 33 parameters associated with $\Omega_X$, $\mu$ and $\Sigma$, along with the remaining cosmological parameters, are fit to several combinations of datasets using {\tt MGCosmoMC}\footnote{\url{https://github.com/sfu-cosmo/MGCosmoMC}}~\cite{Zhao:2008bn,Hojjati:2011ix,Zucca:2019xhg}, which is a modification of {\tt CosmoMC}\footnote{\url{http://cosmologist.info/cosmomc/}}~\cite{Lewis:2002ah}. Additional details on the method, along with a detailed discussion of the Horndeski prior and its role, are presented in SI Sec \ref{sec:method}.

\subsection{Datasets}
 
Our datasets include the CMB temperature, polarization and CMB weak lensing spectra from Planck~\cite{Aghanim:2019ame}, the latest collection of the BAO data from eBOSS~\cite{Alam:2020sor}, MGS~\cite{Ross:2014qpa} and 6dF~\cite{Beutler:2011hx}, the RSD measurements by eBOSS~\cite{Bautista:2020ahg,deMattia:2020fkb, Hou:2020rse,Neveux:2020voa}, the Pantheon SN catalogue~\cite{Scolnic:2017caz} and the DES Year $1$ galaxy clustering and weak lensing data~\cite{Abbott:2017wau} limited to large linear scales~\cite{Zucca:2019xhg}. In addition, for some of the tests, the intrinsic SN magnitude $M_{\rm SN}$ determination by SH0ES~\cite{Riess:2020fzl} was also used, which, when coupled with the Pantheon SN, provides a measurement of $H_0$. Our``Baseline'' dataset includes CMB, BAO and SN, and is combined with DES and RSD to form the ``Baseline + LSS'' dataset used for the reconstructions in Fig.~\ref{fig:all_functions}.

\subsection{Significance of the detection}
\begin{table}[!ht]
\setlength{\tabcolsep}{12pt}
\centering
\begin{tabular}{@{}cccccc@{}}
\toprule
SNR                 & $\Omega_X$ & $\mu$ & $\Sigma$ & $\gamma$\\
\toprule
no theory prior     & & & & \\
\colrule                                                                                                               
Baseline			   & 3.0 & 1.6 & 2.0 & 1.0 \\
Baseline+LSS        & 2.9 & 1.7 & 2.4 & 1.9 \\
All                 & 3.8 & 2.1 & 2.6 & 2.3 \\
Baseline+$A_L$	   & 3.1 & 1.5 & 1.9 & 1.0 \\
Baseline+LSS+$A_L$  & 3.3 & 1.5 & 1.8 & 1.0 \\
All+$A_L$           & 4.0 & 1.9 & 2.2 & 1.3 \\
\toprule
with Horndeski prior & & & & \\
\colrule                                                                                                               
Baseline            & 1.1 & 0.6 & 1.8 & 0.6 \\
Baseline+LSS        & 1.3 & 1.6 & 2.3 & 1.8 \\
All                 & 2.3 & 2.2 & 2.7 & 2.4 \\
Baseline+$A_L$       & 1.2 & 0.5 & 1.7 & 0.5 \\
Baseline+LSS+$A_L$   & 1.7 & 1.3 & 1.3 & 0.9 \\
All+$A_L$           & 2.6 & 2.1 & 1.7 & 1.2 \\
\botrule
\end{tabular}
\caption{\label{tab:snr}
The signal-to-noise ratio (SNR) in the detection of departure of  $\Omega_X$, $\mu$, $\Sigma$ and $\gamma$ from their $\Lambda$CDM values. The gravitational slip $\gamma$ is defined in~(\ref{eq:slip}).
}
\end{table}

Table~\ref{tab:snr} lists the signal-to-noise ratios (SNR) in the detection of departures of  $\Omega_X$, $\mu$ and $\Sigma$ from their $\Lambda$CDM values:
\begin{align} \label{eq:GR_snr}
{\rm SNR}^2 \equiv (\theta-\theta_{\rm GR})^T \mathcal{C}_p^{-1} (\theta-\theta_{\rm GR})
\end{align}
where the vector $\theta$ includes parameters related to the three MG functions, $\mathcal{C}_p$ their covariance and $\theta_{\rm GR}$ represents their GR limit.
The table also shows the SNR for the gravitational slip $\gamma$. 
One can see that the significance of the detection generally drops after including the Horndeski prior, most notably, from $\sim$3$\sigma$ to $\sim$1$\sigma$ for $\Omega_X$ when the SH0ES data is not used. The most persistent deviation is in $\Sigma$, which is above 2$\sigma$ with and without the prior and, to a lesser extent, in $\mu$, because of its strong correlation with $\Sigma$. One can also see that this is largely driven by the CMB lensing anomaly, as the inclusion of $A_L$ as a free parameter brings the SNR in $\Sigma$ down to $\sim$1$\sigma$. Interestingly, the inclusion of the SH0ES prior does not only increase the SNR in $\Omega_X$ but also in $\mu$ and $\Sigma$, due to a non-negligible correlation between the background expansion and the growth rate, as one can also see from Panel (c) of Fig.~\ref{fig:prior_correlations}.

\section{Supplemental Information}

\subsection{Additional information on the formalism, reconstruction method and the theory prior}
\label{sec:method}

\begin{figure*}[tbph!]
\includegraphics[width=0.9\columnwidth]{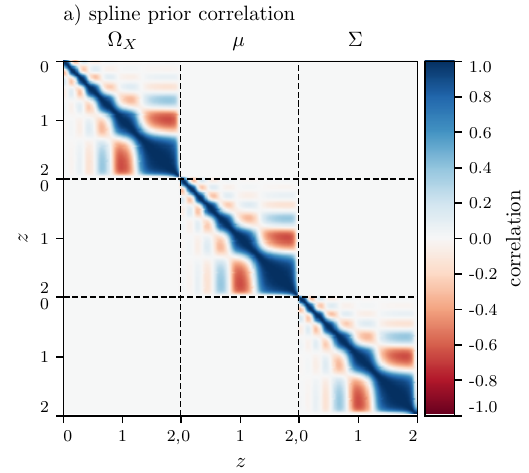}
\includegraphics[width=0.9\columnwidth]{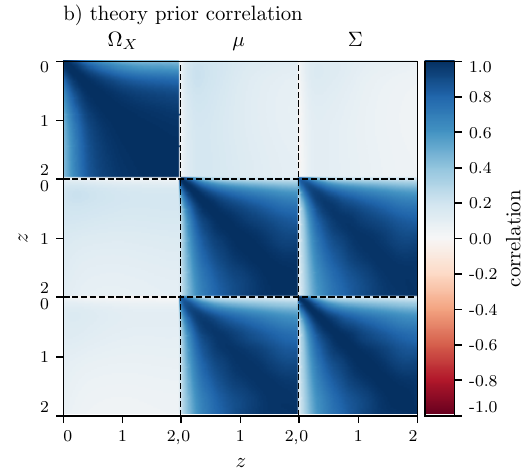}
\includegraphics[width=0.9\columnwidth]{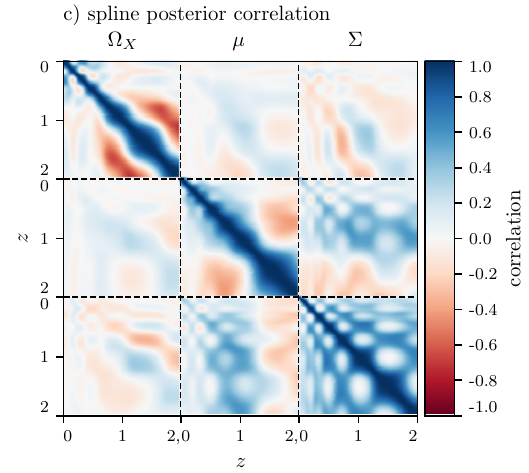}
\includegraphics[width=0.9\columnwidth]{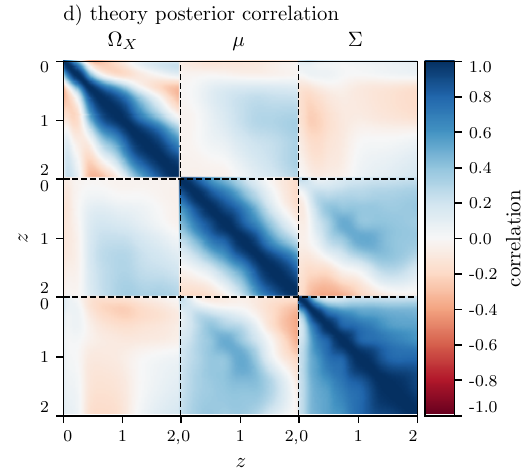}
\caption{ \label{fig:prior_correlations}
a) The implicit correlation prior, as a function of redshift, induced by using the cubic spline to connect the 11 redshift nodes.
All three functions, $\Omega_X$, $\mu$ and $\Sigma$, are subject to the same implicit prior, with no cross-correlation between different functions. b) The Horndeski prior correlating the nodes of $\Omega_X$, $\mu$ and $\Sigma$. The correlation between the nodes of each function is much stronger than that introduced by the cubic spline. The Horndeski prior also introduces a strong correlation between $\mu$ and $\Sigma$. c) The correlation obtained from our ``Baseline'' data posterior covariance of the nodes, {\it i.e.} that determined by the data and the implicit prior correlation in Panel (a). d) The correlation corresponding to the posterior covariance derived from the Baseline data with the help of the Horndeski prior in Panel (b).
}
\end{figure*} 

The starting point of our reconstruction is to parametrize $\Omega_X$, $\mu$ and $\Sigma$ in terms of their values at $11$ discrete values (nodes) of $a$. From the $11$ nodes, $10$ values are distributed uniformly in the interval $a\in[1,0.25]$ (corresponding to $z\in [0,3]$) with another one at $a=0.2$ ($z=4$). We make the functions approach their $\Lambda$CDM values at higher redshifts, because the theoretical prior is obtained by looking at models that deviate from GR at late times only, though studying earlier times deviations from GR is generally possible within the same framework~\cite{Lin:2018nxe}. To allow for a smooth transition between their values at $z=4$ and $z=1000$, we add a set of $9$ anchor nodes arranged along a $\tanh$ pattern, and then use cubic spline to interpolate between all the nodes to obtain continuous functions $\Omega_X(a)$, $\mu(a)$ and $\Sigma(a)$. 
Our results do not depend on how many nodes we use, since, with 10 nodes, we already include many nodes per prior correlation length and additional nodes will be made redundant by the correlation prior.
In addition, the BAO, RSD, DES and SN data probe $z \lesssim 3$, while CMB constrains the integrated effect over $z \lesssim 1000$, with no data in the $3 < z < 1000$ range. 

The cubic spline introduces an implicit smoothness prior into the reconstruction that suppresses sharp changes of the functions between nodes. Panel (a) of Fig.~\ref{fig:prior_correlations} shows the correlation between the nodes of $\Omega_X$, $\mu$ and $\Sigma$ imposed by the cubic spline. As one can see by comparing to Panel (b), this prior is substantially weaker than that derived from the Horndeski theories, as discussed below. Panels (c) and (d) show, respectively, the correlation imposed by data only (which includes the implicit prior), and by data in combination with the Horndeski prior. 

We use an appropriately modified version of {\tt MGCosmoMC}\footnote{\url{https://github.com/sfu-cosmo/MGCosmoMC}}~\cite{Zhao:2008bn,Hojjati:2011ix,Zucca:2019xhg}, based on {\tt CosmoMC}\footnote{\url{http://cosmologist.info/cosmomc/}}~\cite{Lewis:2002ah}, to sample the parameter space, which, in addition to the node parameters $\Omega_{Xi}$, $\mu_i$, $\Sigma_i$ introduced earlier, includes the usual cosmological parameters: $\Omega_bh^2$, $\Omega_ch^2$, $\theta_\star$, $\tau$, $A_s$, $n_s$, $\mathcal{N}$, where $\Omega_{b}h^2$ and $\Omega_{c}h^2$ are the physical densities of baryons and CDM, $\theta_\star$ is the angular size of the sound horizon at the decoupling epoch, $\tau$ is the reionization optical depth, $A_s$ and $n_s$ are the amplitude and the spectral index of primordial fluctuations, and $\mathcal{N}$ collectively denotes the nuisance parameters that appear in various data likelihoods. We note that the last node of $\Omega_X$, corresponding to $a=1$, is not varied as it is the same as the derived parameter $\Omega_{\rm DE}$. We run 8 MCMC chains and assess their convergence through the Gelman-Rubin criterion, assuming the chains have reached convergence when $R-1 \lesssim 0.1$ for the least converged eigenvalue, ensuring that all others have a higher degree of convergence ($R-1 \sim 0.01$).

In addition to performing the reconstruction of $\Omega_X(a)$, $\mu(a)$ and $\Sigma(a)$ by determining the best fit node parameters from data alone, we use the method of~\cite{Crittenden:2011aa,Crittenden:2005wj} to add the Horndeski prior that correlates the nodes $\{\Omega_{Xi},\mu_i,\Sigma_i \} \equiv {\bf f}$. It is introduced as a Gaussian prior
\be
{\cal P}_{\rm prior} \propto \exp [-({\bf f}-{\bf f}_{\rm fid}) {\cal C}^{-1} ({\bf f}-{\bf f}_{\rm fid})^T] \ ,
\ee
where ${\cal C}$ is the correlation matrix derived from the joint covariance of the three functions obtained in~\cite{Espejo:2018hxa}. While we have the full covariance at our disposal, along with the mean values, we opt not to use the latter as our fiducial values ${\bf f}_{\rm fid}$ in order to avoid biasing the outcome of the reconstruction, and also use the normalized correlation matrix for ${\cal C}$. In practice, the prior is implemented as a new contribution to the total $\chi^2$, with ${\bf f}_{\rm fid}$ determined during sampling using the so-called ``running average'' method~\cite{Crittenden:2011aa}. The theory prior acts much like a Wiener filter, discouraging (but not completely prohibiting) abrupt variations of the functions.

Panel (b) of Fig.~\ref{fig:prior_correlations} shows the Horndeski correlation prior used in our work. One can clearly see that the correlation ``length'' is much longer than that of the implicit prior due to the cubic spline shown in Panel (a). This ensures that the prior aided reconstruction is independent of the binning scheme. Also notable is the nearly perfect correlation between $\mu$ and $\Sigma$, in line with the $(\Sigma-1)(\mu-1) \ge 0$ conjecture made in~\cite{Pogosian:2016pwr}. The correlation between $\Omega_X$ and $\mu$ or $\Sigma$ is nearly absent, although, as mentioned above, it can be strong in certain subclasses of Horndeski theories.

We do not consider the $k$-dependence of $\mu$ and $\Sigma$, focusing solely on their evolution with redshift. This helps to reduce the computational costs and is largely justified by the fact that, in practically all known MG theories, the scale-dependence of these functions manifests itself outside the range of linear scales probed by large scale structure (LSS) surveys~\cite{Joyce:2014kja,Wang:2012kj}. Most of the previous cosmological tests of GR have focused on separately testing either the background expansion or modified growth and/or employed simple {\it ad hoc} parametrizations of $w$, $\mu$ and $\Sigma$, or their equivalents~\cite{Planck:2015bue}, which can bias the results and prevent capturing important information in the data, as demonstrated in \cite{recon_prd}. We note that three functions equivalent to $\Omega_X$, $\mu$ and $\Sigma$ were reconstructed in \cite{Pinho:2018unz} from the so-called ``observables'' \cite{Amendola:2012ky}, or model-independent combinations of the data, with a particular focus on constraining the gravitational slip $\gamma$. While very interesting, the methods in that work did not allow determination of the cosmological parameters and addressing the tensions, nor considered the effect of correlations between the functions that could come from theory. A reconstruction of the free functions in the effective theory description of Horndeski \cite{Bloomfield:2012ff,Gleyzes:2013ooa} was performed using a similar method in~\cite{Raveri:2019mxg, Park:2021jmi}, showing interesting hints of departures from GR, albeit at low statistical significance. Our reconstruction has the benefit of not being restricted to Horndeski theories, while still allowing us to check the consistency of various subclasses of Horndeski theories with the data.

We restricted our studies to late-time modifications of $\Lambda$CDM because there one can make a clear connection to the theoretical predictions in the quasi-static limit, and also because extending our phenomenological approach to earlier times would require making additional assumptions about the effect of modified gravity on radiation, which makes the framework less generic. In~\cite{Lin:2018nxe}, a similar phenomenological parametrization was used to study effects of modified gravity at times around recombination, finding that this could help to ease the $H_0$ tensions. Also, in~\cite{Moss:2021obd}, a reconstruction of the dark energy density was performed using methods similar to ours, finding that this can resolve the Hubble tension, but not the $S_8$ tension.

\subsection{Additional information on datasets}
\label{sec:data}
We consider combinations of the following datasets: 
\begin{itemize}
\item ``Planck'': the 2018 release of the Planck CMB temperature, polarization and the reconstructed CMB weak lensing spectra~\cite{Aghanim:2019ame};
\item ``BAO'': the eBOSS  DR16 BAO compilation from~\cite{Alam:2020sor} that includes measurements at multiple redshifts from the samples of Luminous Red Galaxies (LRGs), Emission Line Galaxies (ELGs), clustering quasars (QSOs), and the Lyman-$\alpha$ forest~\cite{Zhao:2020tis,Wang:2020tje,Hou:2020rse,duMasdesBourboux:2020pck}, along with the SDSS DR7 MGS~\cite{Ross:2014qpa} data. We also add the BAO measurement from 6dF~\cite{Beutler:2011hx}. 
This compilation covers the BAO measurements at $0.07 < z < 3.5$. Note that the BAO data considered here are the ``tomographic''  version of the DR12 BOSS BAO at  $0.20 < z < 0.75$ \cite{BOSS:2016lpe} (not the ``consensus'' version using effective redshifts presented in \cite{Alam:2020sor}).
\item ``SN'': the Pantheon SN sample at $0.01 < z < 2.3$~\cite{Scolnic:2017caz};
\item ``RSD'': the eBOSS joint measurement of BAO and RSD for LRGs, ELGs and QSOs~\cite{Bautista:2020ahg,deMattia:2020fkb, Hou:2020rse,Neveux:2020voa}, using it instead of the eBOSS BAO-only measurement. For LRGs, it combines eBOSS LRGs and BOSS CMASS galaxies spanning the redshift range $0.6<z<1$, at an effective redshift of $z_{\rm eff}=0.698$. QSOs cover $0.8<z<2.2$ with an effective redshift of $z_{\rm eff}=1.48$, while ELGs cover $0.6<z<1.1$ with an effective redshift of $z_{\rm eff}=0.845$. In addition, we add BAO-only measurements from 6dF and MGS.
\item ``DES'': the Dark Energy Survey Year 1 measurements of the angular two-point correlation functions of galaxy clustering, cosmic shear and galaxy-galaxy lensing with source galaxies at $0.2< z < 1.3$~\cite{Abbott:2017wau}; since our formalism has no nonlinear prescription for structure formation, the angular separations probing the nonlinear scales were removed using the ``aggressive'' cut option of {\tt MGCAMB} described in~\cite{Zucca:2019xhg}, which uses the method introduced in~\cite{Planck:2015bue,Abbott:2017wau}.
\item ``'M$_{\rm SN}$'': the SH0ES determination of the intrinsic SN type Ia brightness magnitude as obtained by~\cite{Riess:2020fzl}, included in the Pantheon SN likelihood as in~\cite{Benevento:2020fev}. This provides a measurement of $H_0=73.2\pm 1.3$ km/s/Mpc in $\Lambda$CDM. 
\end{itemize}
Our baseline dataset combination (labelled ``Baseline'' from now on) includes Planck, BAO and SN. In addition, we also consider the additional Baseline+RSD+DES and Baseline+RSD+DES+M$_{\rm SN}$. Note that, when RSD is included in the combination, the BAO data do not coincide with the one used in Baseline for the eBOSS LRGs BAO measurement, as we replace it with the joint RSD-BAO measurement. We note that a non-linear modelling of RSD is required to extract the linear growth rate $f\sigma_8$, which was done assuming $\Lambda CDM$. In principle, there can be additional non-linear effects from modified gravity that could bias the $\Lambda CDM$-based extraction of $f\sigma_8$. For specific modified gravity models with the scale independent growth, this bias was shown to be negligible for current measurements~\cite{Barreira:2016ovx}. However, this conclusion depends on the theory of gravity as well as the accuracy of the measurements~\cite{Bose:2017myh}.

For brevity, we refer to RSD+DES as simply ``LSS'', and to Baseline+RSD+DES+M$_{\rm SN}$ as ``All''.

\subsection{Extended analysis of cosmological tensions}
\label{sec:tensions}

\begin{figure*}[tpbh!]
\includegraphics[width=1.8\columnwidth]{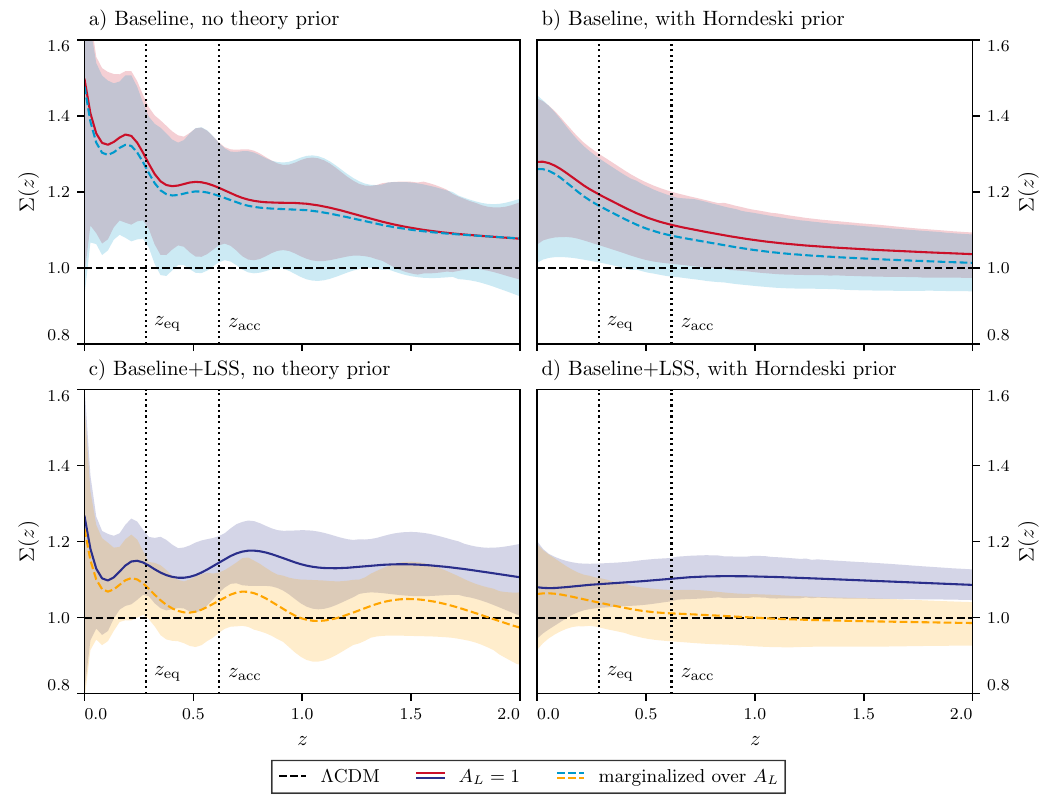}
\caption{Reconstruction of $\Sigma(z)$ from the Baseline (top panels) and the Baseline+LSS (bottom panels) data combinations, without (left panels) and with adding the Horndeski correlation prior (right panels). Solid lines show the reconstructed mean values with the lensing amplitude parameter $A_{\rm L}$ fixed to unity (thus corresponding to the results shown in Fig.~\ref{fig:all_functions}), while the dashed lines correspond to the case where $A_{\rm L}$ was free to vary. The shaded regions show the corresponding $68\%$ confidence levels.}
\label{fig:sigma_al}
\end{figure*}

\begin{figure}[tbph!]
\includegraphics[width=0.9\columnwidth]{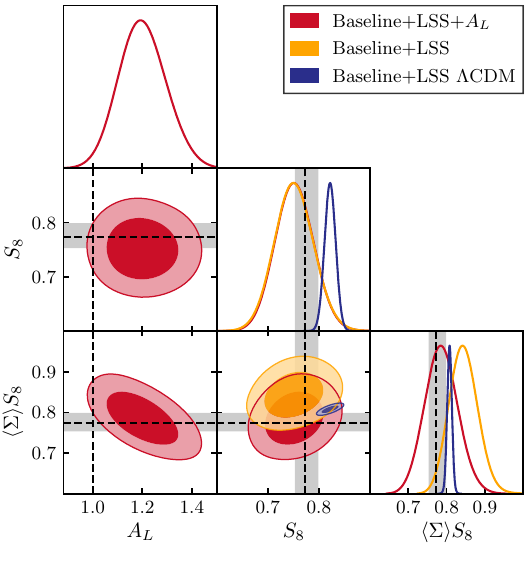}
\caption{ \label{fig:ALandS8}
The $68\%$ and $95\%$ confidence level contours for the parameters $A_{\rm L}$, $S_8$ and the combination $\langle \Sigma \rangle S_8$ derived from the Baseline+LSS data with the Horndeski prior.
Yellow contours refer to the case where $A_{\rm L}=1$, while the red contours represent the case with $A_{\rm L}$ being a free parameters. The blue contours show the $\Lambda$CDM fit to Baseline+LSS. The grey bands show the constraints on $S_8$ and $\langle \Sigma \rangle S_8$ obtained by DES in the $\Lambda$CDM model, in which $\langle \Sigma \rangle=1$. 
}
\end{figure} 

\begin{figure*}[tbph!]
\includegraphics[width=0.9\columnwidth]{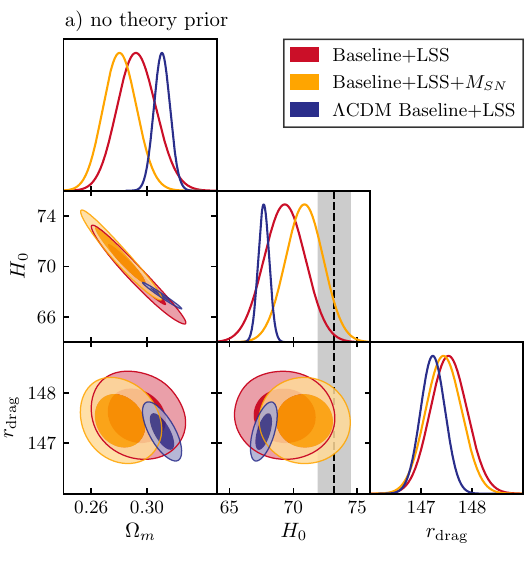}
\includegraphics[width=0.9\columnwidth]{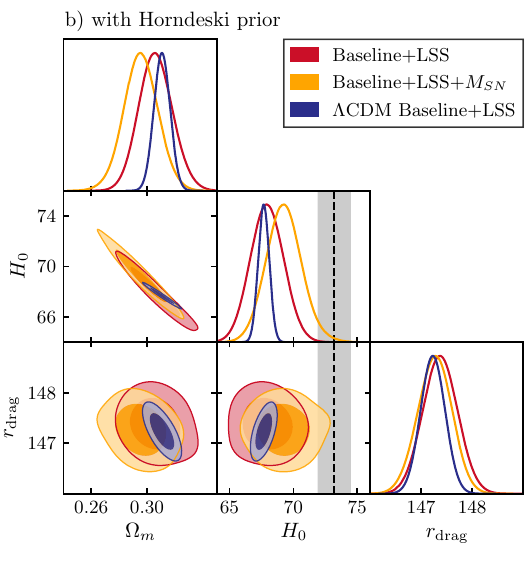}
\caption{ \label{fig:params1}	
$68\%$ and $95\%$ confidence level contours $\Omega_m$, $H_0$ and the sound horizon at baryon decoupling $r_{\rm drag}$, with and without the SH0ES prior  (yellow and red contours, respectively) and with and without the Horndeski prior (left and right panels, respectively). The blue contours show the $\Lambda$CDM fit to Baseline+LSS, while the grey band shows the constrain on $H_0$ obtained by the SH0ES collaboration.
} 
\end{figure*}   

Including the lensing amplitude $A_{\rm L}$~\cite{Calabrese:2008rt} as an extra free parameter helps to assess if the high redshift departure of $\Sigma(z)$ from the GR limit is indeed due to the CMB lensing anomaly. This is a completely phenomenological parameter that rescales the contribution of weak gravitational lensing to the CMB temperature anisotropy spectrum (TT). In a self-consistent cosmological model one should have $A_{\rm L}=1$.
Fig.~\ref{fig:sigma_al} shows the reconstruction of $\Sigma(z)$ with and without the correlation prior, for both the Baseline and Baseline+LSS data combinations, and with the $A_{\rm L}$ parameter both fixed to unity and free to vary. Comparing the fixed and free $A_{\rm L}$ reconstructions shows how the inclusion of $A_{\rm L}$ completely removes the high $z$ departure from GR, with $\Sigma(z)$ now fully consistent with one. 

We further explore the correlations between $\Sigma(z)$ and $A_{\rm L}$ and its implication for the $S_8$ parameter in Fig.~\ref{fig:ALandS8}. As discussed in the main text, the reconstructed shapes of $\Omega_X$ and $\mu$ allow for slightly lower values of $S_8$ than $\Lambda$CDM. However, $S_8$ is also related to the amplitude of the lensing potential, which in our analysis is modulated by $\Sigma$. The parameter that is constrained by DES is $\langle \Sigma \rangle S_8$ where $ \langle \Sigma \rangle$ is an average of $\Sigma$ in the redshift range relevant for DES, and it is equal to one in the $\Lambda$CDM limit. Despite the lowering of $S_8$, this parameter remains the same as that in $\Lambda$CDM when $A_{\rm L}=1$, as it can be seen in the top panels of Tables~\ref{tab:parametersBaseline} and \ref{tab:parametersBaseRSDDES}, where the results for this case are shown. This is due to the enhancement of $\Sigma$ from the GR limit driven by the CMB lensing anomaly as discussed above. As a consequence, if one keeps a fixed $A_{\rm L}=1$, the quality of the fit to DES data is not better than $\Lambda$CDM even when $\Omega_X$, $\mu$ and $\Sigma$ are free to vary (see Table~\ref{tab:parametersBaseRSDDES}). If one instead allows for $A_{\rm L}$ to be free, an anomalous value of this parameter ``solves'' the CMB lensing anomaly, and $\Sigma$ becomes consistent with the GR limit. In this case, the lowering of $S_8$ leads to lower values of $\langle \Sigma \rangle S_8$ and the fit to the DES data is improved compared with $\Lambda$CDM. Table \ref{tab:parametersBaseRSDDES} shows the values of the $\chi^2$ for the different data considered in the analysis. One can see that, with respect to $\Lambda$CDM, the $A_{\rm L}=1$ analysis improves the DES $\chi^2$ by $0.4$ ($1.2$ without the Hordenski prior), while this improvement increases to $2.8$ ($4.7$ without the Hordenski prior) if $A_{\rm L}$ is free. Overall, this analysis reveals that the late time modifications alone are not able to improve the fit to CMB and DES weak lensing simultaneously. 

We now turn our attention to the impact of including the SH0ES prior on the SN magnitude in the data combination analyzed. As it can be seen in Table~\ref{tab:parametersAll}, the main impact of such an addition is an enlargement of the uncertainties and an increase of the estimated mean value of $H_0$. We find $H_0=69.44 \pm 1.30$ km/s/Mpc, which is consistent with the value obtained by SH0ES within $2\sigma$. 
Fig.~\ref{fig:params1} shows $\Omega_m$, $H_0$ and the sound horizon at baryon decoupling $r_{\rm drag}$, with and without the SH0ES prior and with and without the Horndeski prior. It is possible to notice how, for the Baseline+LSS combination, the additional freedom given by the $\Omega_X$, $\mu$ and $\Sigma$ functions only produces an enlargement of the error on $H_0$, with its mean value being the same as in $\Lambda$CDM. When the SH0ES prior is included we obtain instead the increase of the mean value of $H_0$ as well as a slight shift of $r_{\rm drag}$ with respect to $\Lambda$CDM. 

Fig.~\ref{fig:shoes} shows the difference of the luminosity distance prediction with the SH0ES prior from the prediction of the best-fit $\Lambda$CDM without the SH0ES prior. The Pantheon SN data points are calibrated using the SH0ES measurement of $M_{\rm SN}$ while the BAO data points are converted to the luminosity distance from the angular diameter distance using $d_L=(1+z)^2 d_A$. There are two main issues. The first problem is that the luminosity distance calibrated from CMB in $\Lambda$CDM does not agree with the SN data while it agrees with the BAO data. The second problem is the discrepancy between the BAO and SN data. The latter makes it hard for late modifications to resolve this tension fully as it is not possible to fit BAO and SN data simultaneously unless we change $r_{\rm drag}$ by an early time modification. This is also the case in our reconstruction. Due to the freedom in $\Omega_X(z)$, the luminosity distance at lower redshifts becomes closer to the SN data compared with $\Lambda$CDM, which leads to a larger $H_0$. However, it is still not possible to reproduce the luminosity distance calibrated by the SH0ES measurement of $M_{\rm SN}$ fully. 

\begin{figure*}[tbph!]
\includegraphics[width=0.9\columnwidth]{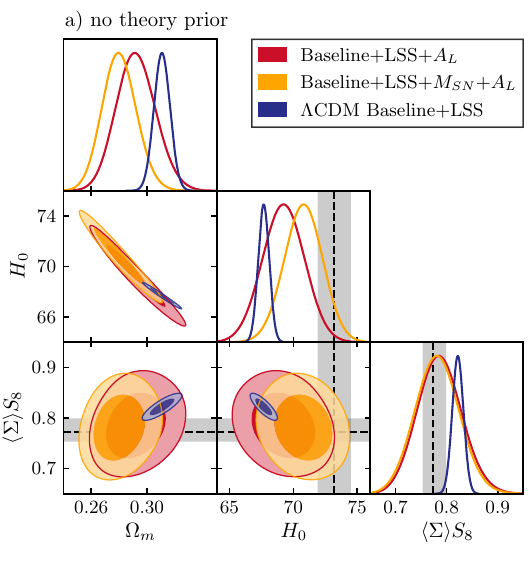}
\includegraphics[width=0.9\columnwidth]{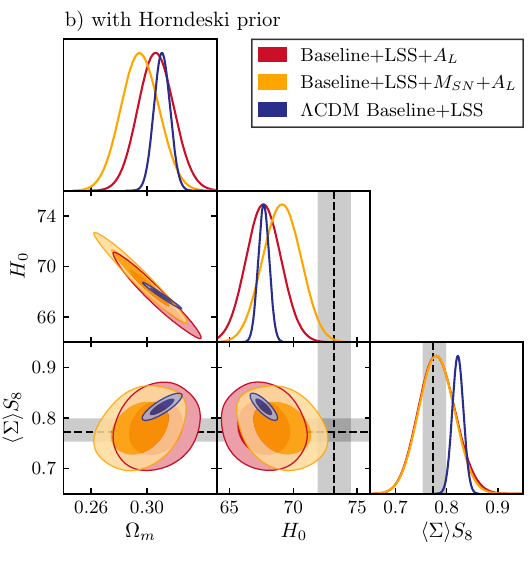}
\caption{ \label{fig:params2}
The $68\%$ and $95\%$ confidence level contours for the tension parameters, $S_8$, $H_0$ and $\Omega_M$, from the Baseline+LSS data with and without the SH0ES prior on the SN magnitude M$_{\rm SN}$ (yellow and red contours, respectively), and with and without the Horndeski prior (left and right panels, respectively). The blue contours show the results obtained in the 
$\Lambda$CDM limit, while the grey band shows the $H_0$ measured by the SH0ES collaboration.
}			
\end{figure*}

To show the extent to which the well known $H_0$ and $S_8$ tensions can be resolved by allowing for time-dependent $\Omega_X$, $\mu$ and $\Sigma$, we show in Fig.~\ref{fig:params2} the constraints on $\Omega_m$, $H_0$ and $\langle \Sigma \rangle S_8$ and compare the results obtained with and without the SH0ES prior. The results shown in this figure refer to the case where $A_{\rm L}$ is considered as a free parameter. As discussed before, without this parameter, the CMB lensing would prevents us from addressing the $S_8$ tension, thus achieving a better fit to DES. On the other hand, allowing for a varying $A_L$ removes the CMB anomaly, making it possible to fit simultaneously to the SH0ES and DES data better than $\Lambda$CDM. As shown in Table~\ref{tab:parametersAll}, the total improvement of $\chi^2$ is $\Delta\chi^2=-25.8$ without the Horndeski prior and $\Delta\chi^2=-14.1$ with the prior.

\subsection{Extended discussion of implications for Horndeski} 
\label{sec:horndeski}

\begin{figure*}[tbph!]
		\includegraphics[width=1.8\columnwidth]{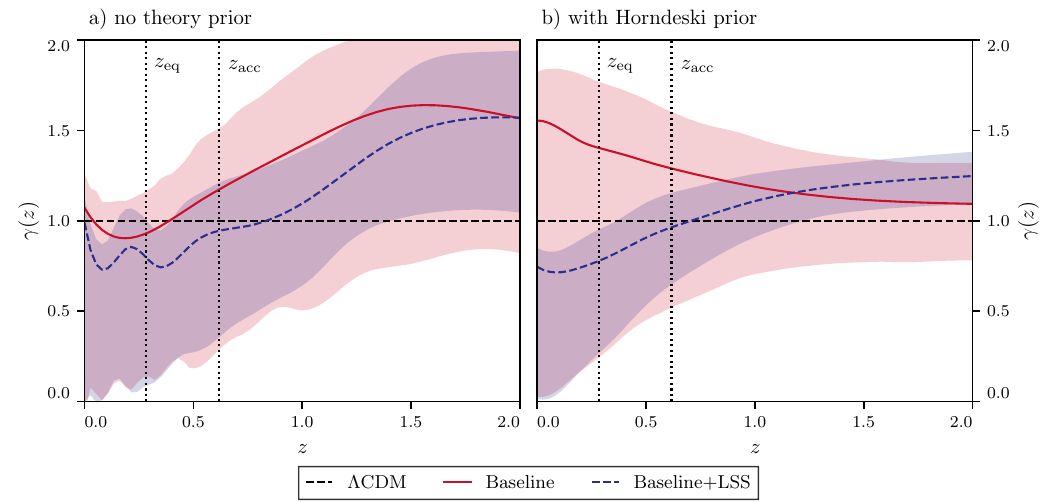}
		\caption{Reconstruction of the gravitational slip $\gamma(z)$ constructed from $\mu(z)$ and $\Sigma(z)$. The reconstruction is obtained using only data (left panel) and adding the Horndeski correlation prior (right panel). Red lines show the mean reconstructions using the Baseline data combination, with the red region showing the $68\%$ confidence level, while blue lines and regions refer to the case where RSD and DES are added to the baseline.	
		}\label{fig:slip}
\end{figure*}  

Any statistically significant departure of $X(z)$, $\mu(z)$ or $\Sigma(z)$ from unity would imply either a break down of the $\Lambda$CDM model or a problem, {\it e.g.} a systematic effect, with some of the datasets. The deviations from the $\Lambda$CDM prediction seen in our reconstructions are at a level of $\sim$2$\sigma$, comparable to the significance of the $S_8$ and $A_L$ tensions. While this hardly prompts an urgent revision of $\Lambda$CDM, it is nevertheless interesting to ask what implications the trends exhibited by the reconstructions would have for alternative models. In what follows, we first interpret the reconstruction results in the context of Horndeski theories, showing that the reconstructed evolution of the three functions, had it been detected at a higher confidence level, would rule out broad classes of scalar-tensor theories.

As shown in~\cite{Silvestri:2013ne}, under the QSA, the expressions for $\mu$ and $\Sigma$ in local theories of gravity take the form of ratios of polynomials in $k$. In Horndeski theories, with second order equations of motion, the polynomials are quadratic in $k$. The scale dependence of $\mu$ and $\Sigma$, however, is unlikely to manifest itself in the range of $k$ probed by large scale structure surveys for which linear perturbation theory is valid. The $k$-dependence marks the transition from the $k \ll \lambda_f^{-1}$ limit, where perturbations of the scalar field can be neglected, to the $k \gg \lambda_f^{-1}$ regime in which the scalar field perturbations mediate a fifth force. The known screening mechanisms in Horndeski theories place the range of $k$ probed by our reconstruction in one of these two limits~\cite{Joyce:2014kja,Wang:2012kj}.

While the QSA is not guaranteed to be accurate in all circumstances, especially on near-horizon scales, it has been found to work quite well for identifying the key phenomenological signatures of Horndeski theories~\cite{Peirone:2017ywi}. Even though we cannot be certain which regime, $k\ll \lambda_f^{-1}$ or $k \gg \lambda_f^{-1}$, is being probed by our scale-independent parametrization, we can still make several useful deduction by comparing our reconstructions to the QSA expressions for $\mu$ and $\Sigma$ in the two limits.

In the $k \ll \lambda_f^{-1}$ limit, the QSA expressions for our functions are~\cite{Pogosian:2016pwr}
\ba
&&\mu \rightarrow \mu_0 = \frac{m_{0}^{2}}{M_{\star}^{2}}c_T^2
\\
&&\Sigma \rightarrow \Sigma_0 = \frac{m_{0}^{2}}{M_{\star}^{2}} \frac{1+c_{T}^{2}}{2}
\\
&&\gamma \rightarrow \gamma_0 = c_T^{-2} \ ,
\label{eq:gamma_0}
\ea
while, for $k \gg \lambda_f^{-1}$, one has \cite{Gleyzes:2015rua,Pogosian:2016pwr}
\ba
&&\mu \rightarrow \mu_\infty = \frac{m_{0}^{2}}{M_{\star}^{2}} {(c_T^2 + \beta_\xi^2)}
\\
&&\Sigma \rightarrow \Sigma_\infty = \frac{m_{0}^{2}}{M_{\star}^{2}} \left( {1+c_{T}^{2} \over 2} + \frac{\beta_\xi^2 + \beta_B \beta_\xi}{2} \right)
\\
&&\gamma \rightarrow \gamma_\infty = {1+\beta_B \beta_\xi \over c_T^2 + \beta_\xi^2} \ ,
\label{eq:gamma_inf}
\ea
where $m_0$ is the Planck mass, 
$M_\star$ is the modified Planck mass, $c_T$ is the propagation speed of tensor metric modes, {\it i.e.} the speed of gravitational waves, while $\beta_B$ and $\beta_\xi$ are two functions, originating from different ways of coupling the metric and the scalar field, that represent the fifth force contribution.

The limiting expressions above highlight the close relationship between the gravitational slip and the tensor speed $c_T$~\cite{Saltas:2014dha}. In particular, $\gamma$ must approach $1$ in the large scale limit if $c_T=1$~\cite{Pogosian:2016pwr}. In the small scale limit, on the other hand, $\gamma \ne 1$ could be either due to a fifth force or $c_T \ne 1$, or both. Since $c_T \ne 1$ is disfavoured by the multi-messenger observations of gravitational waves from binary neutron stars at $z \sim 0$~\cite{TheLIGOScientific:2017qsa,Monitor:2017mdv}, one would conclude that there is evidence for a fifth force at low redshifts.

The subclass of the Horndeski theories with $\gamma=1$, other than $\Lambda$CDM, include the Cubic Galileon~\cite{Deffayet:2009wt}  and KGB~\cite{Deffayet:2010qz} models, along with the so-called ``no-slip gravity''~\cite{Linder:2018jil}, for which $c_T=1, \beta_{\xi}=0$ so that 
\ba
\mu_0 = \Sigma_0 = \mu_\infty = \Sigma_\infty  = \frac{m_{0}^{2}}{M_{*}^{2}}. 
\ea

Fig.~\ref{fig:slip} shows the gravitational slip $\gamma(z)$ derived from the reconstructions of $\mu(z)$ and $\Sigma(z)$. It is important to note that the Baseline dataset is not capable of breaking the degeneracy between $\mu$ and $\Sigma$. Correspondingly, $\gamma(z)$ reconstructed from the Baseline data has a large uncertainty and is strongly prior dependent. Using Baseline+RSD+DES, on the other hand, allows for the degeneracy between $\mu(z)$ and $\Sigma(z)$ to be partially broken. As a result, $\gamma(z)$ is better constrained and the trends in its time-evolution are essentially the same with and without the Horndeski prior. In both cases, one finds $\gamma>1$ at higher $z$ and $\gamma<1$ at lower $z$, with the transition between these two limits happening around the redshift at which cosmic acceleration sets in. In the case with the Horndeski prior, the uncertainties are reduced, making the trend significant at more than $2\sigma$.

Keeping in mind that the significance of the $\gamma \ne 1$ detection is relatively low, one could ask what such a time-dependence would imply for Horndeski theories. Aside from ruling out models with $\gamma=1$, like the no-slip gravity, Cubic Galileons and KGB, the fact that we observe $\gamma > 1$ would rule out the generalized Brans-Dicke models (GBD), which predicts $c_T=1$ and $\beta_{B} = - \beta_{\xi}$, {\it i.e.} all models with a canonical form of the scalar field kinetic energy term. The latter conclusion follows from the fact that one should have $\gamma \le 1$ in GBD on all scales.

\begin{figure*}[tbph!]
\includegraphics[width=1.8\columnwidth]{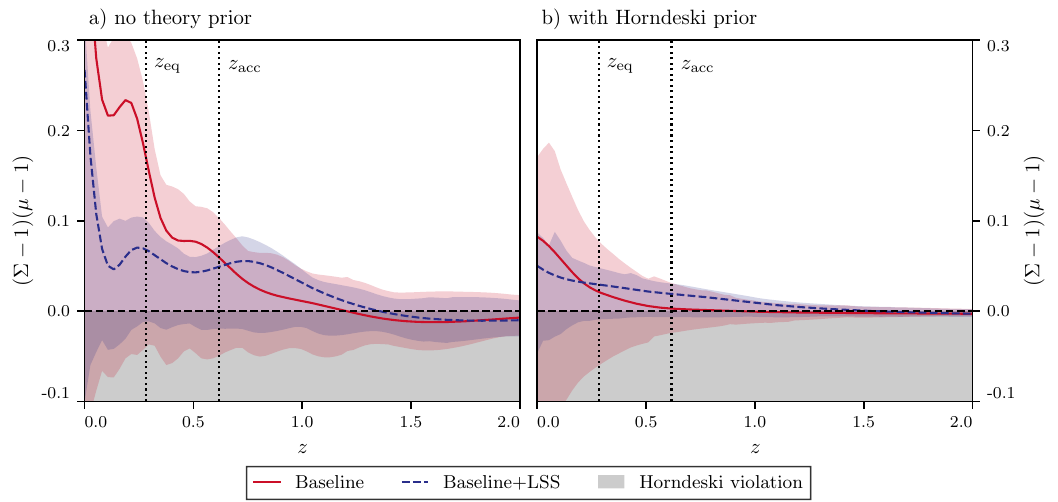}
\caption{ \label{fig:horndeski_consistency}
Reconstruction of the combination $(\Sigma(z)-1)(\mu(z)-1)$, with red lines/regions corresponding to the mean/$68\%$ confidence level for the Baseline data combination, and blue lines/regions representing the case where RSD and DES are added to the Baseline. The gray region shows the area where the Horndeski conjecture is violated.
}
\end{figure*} 

Finally, as pointed out in~\cite{Pogosian:2016pwr} based on analytical considerations in the QSA limit, and later confirmed by a numerical sampling of Horndeski solutions~\cite{Peirone:2017ywi,Espejo:2018hxa}, one expects a strong correlation between $\mu$ and $\Sigma$, with $(\Sigma-1)(\mu-1) \ge 0$. To violate the latter condition, independent sectors/terms of the Horndeski theory would need to conspire to evolve in just the right way for no apparent reason. Thus, looking for signs of violation of the $(\Sigma-1)(\mu-1) \ge 0$ conjecture is an important test of Horndeski. Fig.~\ref{fig:horndeski_consistency} shows the evolution of $(\Sigma-1)(\mu-1)$ derived from our reconstructions. With or without the Horndeski prior, the reconstructions show a good consistency with $(\Sigma-1)(\mu-1) \ge 0$.

\subsection{Parameter tables}

\begin{table*}[!ht]
	\setlength{\tabcolsep}{12pt}
	\centering
	\begin{tabular}{@{}cccccc@{}}
		\toprule
		{\it Baseline} data           & $\Lambda$CDM (reference)             & No theory prior             & With Horndeski prior \\
		\toprule                                                                                                               
		$H_0$                         & $(67.44)\, 67.43^{+0.78}_{-0.81}$    & $(69.86)\, 69.2\pm 1.4$     & $(67.70)\, 67.6^{+1.4}_{-1.2}$       \\
		$\Omega_m$                    & $(0.3141)\, 0.314^{+0.011}_{-0.011}$ & $(0.2890)\, 0.295\pm 0.012$ & $(0.3108)\, 0.310^{+0.010}_{-0.014}$ \\
		$\langle \Sigma \rangle S_8$  & $(0.8319)\, 0.831^{+0.021}_{-0.020}$ & $(0.8529)\, 0.893\pm 0.050$ & $(0.8555)\, 0.901\pm 0.044$           \\
		\colrule
		$\chi^2_{\rm CMB}$            & $2765.7$                             & $2760.7\, (-5)$              & $2761.92\, (-3.8)$ \\
		$\chi^2_{\rm CMBL}$           & $8.7$                                & $9.9\, (+1.2)$               & $9.58\, (+0.9)$ \\
		$\chi^2_{\rm SN}$             & $1035.2$                             & $1030.9\, (-4.3)$            & $1035.5\, (+0.3)$ \\
		$\chi^2_{\rm BAO}$            & $20.6$                               & $15.3\, (-5.3)$              & $21.08\, (+0.5)$ \\
		\colrule                                                                                                                
		$\chi^2_{\rm tot}$            & $3830.2$                             & $3816.9\, (-13.3)$           & $3828.1\, (-2.1)$\\
		\botrule
		                              & $\Lambda$CDM $+A_L$           & No theory prior $+A_L$       & With Horndeski prior $+A_L$ \\
		\toprule                             
		$H_0$                         & $(67.896)\, 67.83\pm 0.46$    & $(69.78)\, 69.3\pm 1.4$              & $(67.15)\, 67.8^{+1.1}_{-1.4}$       \\
		$\Omega_m$                    & $(0.3079)\, 0.3088\pm 0.0061$ & $(0.2902)\, 0.294^{+0.011}_{-0.013}$ & $(0.3122)\, 0.308\pm 0.012$ \\
		$\langle \Sigma \rangle S_8$  & $(0.8114)\, 0.814\pm 0.013$   & $(0.800)\, 0.883^{+0.067}_{-0.081}$  & $(0.817)\, 0.870\pm 0.070$           \\
		$A_L$                         & $(1.0739)\, 1.069\pm 0.036$   & $(1.084)\, 1.027^{+0.093}_{-0.11}$   & $(1.117)\, 1.060^{+0.093}_{-0.13}$ \\
		\colrule
		$\chi^2_{\rm CMB}$            & $2760.2\, (-5.5)$    & $2759.9\, (-5.8)$                        & $2759.6\, (6.1)$ \\
		$\chi^2_{\rm CMBL}$           & $10.0\, (+1.3)$      & $9.7\, (+1.0)$                           & $9.7\, (+1.0)$ \\
		$\chi^2_{\rm SN}$             & $1034.9\, (-0.3)$    & $1030.8\, (-4.4)$                        & $1035.7\, (+0.5)$ \\
		$\chi^2_{\rm BAO}$            & $21.0\, (+0.4)$      & $15.7\, (-4.9)$                          & $20.1\, (-0.5)$ \\
		\colrule                                                                                                                
		$\chi^2_{\rm tot}$            & $3826.1\, (-4.1)$    & $3816.2\, (-14.0)$                        & $3825.1\, (-5.1)$\\
		\botrule
	\end{tabular}
	\caption{ \label{tab:parametersBaseline}
	The best fitting maximum posterior model parameters, in parentheses, mean and $68\%$ C.L. constraints for the {\it Baseline} data combination.
	Data likelihoods at maximum posterior are compared to their reference $\Lambda$CDM values in parenthesis.
}
\end{table*}

\begin{table*}[!ht]
	\setlength{\tabcolsep}{12pt}
	\centering
	\begin{tabular}{@{}cccccc@{}}
		\toprule
		{\it Baseline} + RSD + DES    & $\Lambda$CDM (reference)      & No theory prior             & With Horndeski prior \\
		\toprule                                                                                                               
		$H_0$                         & $(67.632)\, 67.68\pm 0.41$    & $(70.13)\, 69.3\pm 1.5$     & $(68.23)\, 67.9\pm 1.2$       \\
		$\Omega_m$                    & $(0.3115)\, 0.3107\pm 0.0054$ & $(0.2877)\, 0.293^{+0.012}_{-0.014}$ & $(0.3040)\, 0.306^{+0.010}_{-0.012}$ \\
		$\langle \Sigma \rangle S_8$  & $(0.8255)\, 0.822\pm 0.010$   & $(0.8236)\, 0.844^{+0.032}_{-0.037}$ & $(0.8247)\, 0.838\pm 0.025$           \\
		\colrule
		$\chi^2_{\rm CMB}$            & $2766.3$                      & $2761.3\, (-5.0)$              & $2764.4\, (-1.9)$ \\
		$\chi^2_{\rm CMBL}$           & $8.8$                         & $9.3\, (+0.5)$                 & $8.9\, (+0.1)$ \\
		$\chi^2_{\rm SN}$             & $1035.0$                      & $1029.6\, (-5.4)$              & $1034.6\, (-0.4)$ \\
		$\chi^2_{\rm BAO}$            & $19.3$                        & $13.8\, (-5.5)$                & $17.9\, (-1.4)$ \\
		$\chi^2_{\rm DES}$            & $322.5$                       & $321.3\, (-1.2)$               & $322.1\, (-0.4)$ \\
		\colrule                                                                                                                
		$\chi^2_{\rm tot}$            & $4151.8$                      & $4135.3\, (-16.5)$             & $4147.9\, (-3.9)$\\
		\botrule
		                              & $\Lambda$CDM $+A_L$           & No theory prior $+A_L$              & With Horndeski prior $+A_L$ \\
		\toprule                                     
		$H_0$                         & $(68.028)\, 68.20\pm 0.46$    & $(69.64)\, 69.2\pm 1.5$              & $(67.64)\, 67.7\pm 1.3$       \\
		$\Omega_m$                    & $(0.3061)\, 0.3038\pm 0.0059$ & $(0.2898)\, 0.292^{+0.012}_{-0.014}$ & $(0.3071)\, 0.307\pm 0.012$ \\
		$\langle \Sigma \rangle S_8$  & $(0.8084)\, 0.800\pm 0.013$   & $(0.7672)\, 0.786\pm 0.040$          & $(0.7765)\, 0.781^{+0.031}_{-0.036}$           \\
		$A_L$                         & $(1.0777)\, 1.093^{+0.034}_{-0.038}$    & $(1.219)\, 1.199^{+0.082}_{-0.096}$  & $(1.248)\, 1.208\pm 0.090$ \\
		\colrule
		$\chi^2_{\rm CMB}$            & $2760.3\, (-6.0)$  & $2760.5\, (-5.8)$    & $2761.8\, (-4.5)$ \\
		$\chi^2_{\rm CMBL}$           & $10.0\, (+1.2)$    & $10.1\, (+1.3)$      & $10.2\, (+1.4)$ \\
		$\chi^2_{\rm SN}$             & $1034.8\, (-0.2)$  & $1030.2\, (-4.8)$    & $1034.6\, (-0.4)$ \\
		$\chi^2_{\rm BAO}$            & $19.3\, (0.0)$     & $12.4\, (-6.9)$      & $15.1\, (-4.2)$ \\
		$\chi^2_{\rm DES}$            & $321.2\, (-1.3)$   & $317.8\, (-4.7)$     & $319.7\, (-2.8)$ \\
		\colrule                                                                                                                
		$\chi^2_{\rm tot}$            & $4145.6\, (-6.2)$  & $4131.0\, (-20.8)$   & $4141.4\, (-10.4)$\\
		\botrule
	\end{tabular}
	\caption{ \label{tab:parametersBaseRSDDES}
	The best fitting maximum posterior model parameters, in parentheses, mean and $68\%$ C.L. constraints for the {\it Baseline} + RSD + DES data combination.
	Data likelihoods at maximum posterior are compared to their reference $\Lambda$CDM values in parenthesis.
}
\end{table*}

\begin{table*}[!ht]
	\setlength{\tabcolsep}{12pt}
	\centering
	\begin{tabular}{@{}cccccc@{}}
		\toprule
		{\it Baseline} + RSD + DES +$M_{SN}$ & $\Lambda$CDM (reference)      & No theory prior              & With Horndeski prior \\
		\toprule                                                                                                               
		$H_0$                         & $(68.039)\, 68.13\pm 0.40$    & $(71.34)\, 70.9\pm 1.4$              & $(69.44)\, 69.2\pm 1.3$     \\
		$\Omega_m$                    & $(0.3062)\, 0.3050\pm 0.0052$ & $(0.2759)\, 0.281\pm 0.011$          & $(0.2942)\, 0.296\pm 0.012$ \\
		$\langle \Sigma \rangle S_8$  & $(0.8107)\, 0.8134\pm 0.0099$ & $(0.8192)\, 0.836^{+0.031}_{-0.037}$ & $(0.8365)\, 0.830\pm 0.025$ \\
		\colrule
		$\chi^2_{\rm CMB}$            & $2768.4$ & $2763.8\, (-4.6)$  & $2764.0\, (-4.4)$ \\
		$\chi^2_{\rm CMBL}$           & $10.1$   & $8.8\, (-1.3)$     & $9.0\, (-1.1)$ \\
		$\chi^2_{\rm SN}$             & $1025.9$ & $1022.9\, (-3.0)$  & $1025.8\, (-0.1)$ \\
		$\chi^2_{\rm BAO}$            & $19.5$   & $15.5\, (-4.0)$    & $20.6\, (+1.1)$ \\
		$\chi^2_{\rm DES}$            & $321.2$  & $319.2\, (-2.0)$   & $321.5\, (+0.3)$ \\
		$\chi^2_{\rm M}$              & $18.8$   & $11.9\, (-6.9)$    & $12.9\, (-5.9)$ \\
		\colrule                                                                                                                
		$\chi^2_{\rm tot}$            & $4163.9$ & $4142.1\, (-21.8)$ & $4153.8\, (-10.1)$\\
		\botrule
		                              & $\Lambda$CDM $+A_L$           & No theory prior $+A_L$              & With Horndeski prior $+A_L$ \\
		\toprule                                     
		$H_0$                         & $(68.566)\, 68.71\pm 0.44$    & $(70.96)\, 70.8\pm 1.4$              & $(69.67)\, 69.1\pm 1.4$      \\
		$\Omega_m$                    & $(0.2992)\, 0.2974\pm 0.0056$ & $(0.2780)\, 0.280^{+0.011}_{-0.012}$ & $(0.2935)\, 0.295\pm 0.013$  \\
		$\langle \Sigma \rangle S_8$  & $(0.7922)\, 0.789\pm 0.013$   & $(0.7584)\, 0.782\pm 0.040$          & $(0.7802)\, 0.780\pm 0.033$         \\
		$A_L$                         & $(1.1067)\, 1.110\pm 0.037$   & $(1.209)\, 1.190^{+0.080}_{-0.096}$  & $(1.118)\, 1.190^{+0.085}_{-0.099}$ \\
		\colrule
		$\chi^2_{\rm CMB}$            & $2760.7\, (-7.7)$  & $2762.1\, (-6.3)$    & $2762.1\, (-6.3)$ \\
		$\chi^2_{\rm CMBL}$           & $10.1\, (0.0)$     & $9.8\, (-0.3)$       & $10.0\, (-0.1)$ \\
		$\chi^2_{\rm SN}$             & $1025.8\, (-0.1)$  & $1021.6\, (-4.3)$    & $1025.7\, (-0.2)$ \\
		$\chi^2_{\rm BAO}$            & $20.7\, (+1.2)$    & $14.6\, (-4.9)$      & $19.8\, (+0.3)$ \\
		$\chi^2_{\rm DES}$            & $320.2\, (-1.0)$   & $317.6\, (-3.6)$     & $319.2\, (-2.0)$ \\
		$\chi^2_{\rm M}$              & $15.8\, (-3.0)$    & $12.4\, (-6.4)$      & $13.0\, (-5.8)$ \\
		\colrule                                                                                                                
		$\chi^2_{\rm tot}$            & $4153.3\, (-10.6)$ & $4138.1\, (-25.8)$   & $4149.8\, (-14.1)$\\
		\botrule
	\end{tabular}
	\caption{ \label{tab:parametersAll}
	The best fitting maximum posterior model parameters, in parentheses, mean and $68\%$ C.L. constraints for the {\it Baseline} + RSD + DES + $M_{SN}$ data combination. Data likelihoods at maximum posterior are compared to their reference $\Lambda$CDM values in parenthesis.
	}
\end{table*}

\acknowledgments 
For the purpose of open access, the author(s) has applied a Creative Commons Attribution (CC BY) licence to any Author Accepted Manuscript version arising. 
LP is supported by the National Sciences and Engineering Research Council (NSERC) of Canada, and by the Chinese Academy of Sciences President's International Fellowship Initiative, Grant No. 2020VMA0020. 
MR is supported in part by NASA ATP Grant No. NNH17ZDA001N and by funds provided by the Center for Particle Cosmology.
KK was supported by the European Research Council under the European Union's Horizon 2020 programme (grant agreement No.646702 ``CosTesGrav"). KK is supported the UK STFC grant ST/S000550/1 and ST/W001225/1. MM has received the support of a fellowship from `la Caixa' Foundation (ID 100010434), with fellowship code LCF/BQ/PI19/11690015, and the support of the Spanish Agencia Estatal de Investigacion through the grant `IFT Centro de Excelencia Severo Ochoa SEV-2016-0599'. AS acknowledges support from the NWO and the Dutch Ministry of Education, Culture and Science (OCW). GBZ is supported by the National Key Basic Research and Development Program of China (No. 2018YFA0404503), NSFC Grants 11925303, 11720101004, 11890691, a grant of CAS Interdisciplinary Innovation Team, and science research grants from the China Manned Space Project with NO. CMS-CSST-2021-B01. We gratefully acknowledge using {\tt GetDist} \cite{Lewis:2019xzd}. This research was enabled in part by support provided by WestGrid ({\tt www.westgrid.ca}), Compute Canada Calcul Canada ({\tt www.computecanada.ca}) and by the University of Chicago Research Computing Center through the Kavli Institute for Cosmological Physics at the University of Chicago.
Supporting research data are available on reasonable request from the corresponding author.

\bibliography{mgrecon}

\begin{thebibliography}{81}%
\makeatletter
\providecommand \@ifxundefined [1]{%
 \@ifx{#1\undefined}
}%
\providecommand \@ifnum [1]{%
 \ifnum #1\expandafter \@firstoftwo
 \else \expandafter \@secondoftwo
 \fi
}%
\providecommand \@ifx [1]{%
 \ifx #1\expandafter \@firstoftwo
 \else \expandafter \@secondoftwo
 \fi
}%
\providecommand \natexlab [1]{#1}%
\providecommand \enquote  [1]{``#1''}%
\providecommand \bibnamefont  [1]{#1}%
\providecommand \bibfnamefont [1]{#1}%
\providecommand \citenamefont [1]{#1}%
\providecommand \href@noop [0]{\@secondoftwo}%
\providecommand \href [0]{\begingroup \@sanitize@url \@href}%
\providecommand \@href[1]{\@@startlink{#1}\@@href}%
\providecommand \@@href[1]{\endgroup#1\@@endlink}%
\providecommand \@sanitize@url [0]{\catcode `\\12\catcode `\$12\catcode
  `\&12\catcode `\#12\catcode `\^12\catcode `\_12\catcode `\%12\relax}%
\providecommand \@@startlink[1]{}%
\providecommand \@@endlink[0]{}%
\providecommand \url  [0]{\begingroup\@sanitize@url \@url }%
\providecommand \@url [1]{\endgroup\@href {#1}{\urlprefix }}%
\providecommand \urlprefix  [0]{URL }%
\providecommand \Eprint [0]{\href }%
\providecommand \doibase [0]{http://dx.doi.org/}%
\providecommand \selectlanguage [0]{\@gobble}%
\providecommand \bibinfo  [0]{\@secondoftwo}%
\providecommand \bibfield  [0]{\@secondoftwo}%
\providecommand \translation [1]{[#1]}%
\providecommand \BibitemOpen [0]{}%
\providecommand \bibitemStop [0]{}%
\providecommand \bibitemNoStop [0]{.\EOS\space}%
\providecommand \EOS [0]{\spacefactor3000\relax}%
\providecommand \BibitemShut  [1]{\csname bibitem#1\endcsname}%
\let\auto@bib@innerbib\@empty
\bibitem [{\citenamefont {Silvestri}\ and\ \citenamefont
  {Trodden}(2009)}]{Silvestri:2009hh}%
  \BibitemOpen
  \bibfield  {author} {\bibinfo {author} {\bibfnamefont {Alessandra}\
  \bibnamefont {Silvestri}}\ and\ \bibinfo {author} {\bibfnamefont {Mark}\
  \bibnamefont {Trodden}},\ }\bibfield  {title} {\enquote {\bibinfo {title}
  {{Approaches to Understanding Cosmic Acceleration}},}\ }\href {\doibase
  10.1088/0034-4885/72/9/096901} {\bibfield  {journal} {\bibinfo  {journal}
  {Rept. Prog. Phys.}\ }\textbf {\bibinfo {volume} {72}},\ \bibinfo {pages}
  {096901} (\bibinfo {year} {2009})},\ \Eprint {http://arxiv.org/abs/0904.0024}
  {arXiv:0904.0024 [astro-ph.CO]} \BibitemShut {NoStop}%
\bibitem [{\citenamefont {Joyce}\ \emph {et~al.}(2015)\citenamefont {Joyce},
  \citenamefont {Jain}, \citenamefont {Khoury},\ and\ \citenamefont
  {Trodden}}]{Joyce:2014kja}%
  \BibitemOpen
  \bibfield  {author} {\bibinfo {author} {\bibfnamefont {Austin}\ \bibnamefont
  {Joyce}}, \bibinfo {author} {\bibfnamefont {Bhuvnesh}\ \bibnamefont {Jain}},
  \bibinfo {author} {\bibfnamefont {Justin}\ \bibnamefont {Khoury}}, \ and\
  \bibinfo {author} {\bibfnamefont {Mark}\ \bibnamefont {Trodden}},\ }\bibfield
   {title} {\enquote {\bibinfo {title} {{Beyond the Cosmological Standard
  Model}},}\ }\href {\doibase 10.1016/j.physrep.2014.12.002} {\bibfield
  {journal} {\bibinfo  {journal} {Phys. Rept.}\ }\textbf {\bibinfo {volume}
  {568}},\ \bibinfo {pages} {1--98} (\bibinfo {year} {2015})},\ \Eprint
  {http://arxiv.org/abs/1407.0059} {arXiv:1407.0059 [astro-ph.CO]} \BibitemShut
  {NoStop}%
\bibitem [{\citenamefont {Koyama}(2016)}]{Koyama:2015vza}%
  \BibitemOpen
  \bibfield  {author} {\bibinfo {author} {\bibfnamefont {Kazuya}\ \bibnamefont
  {Koyama}},\ }\bibfield  {title} {\enquote {\bibinfo {title} {{Cosmological
  Tests of Modified Gravity}},}\ }\href {\doibase
  10.1088/0034-4885/79/4/046902} {\bibfield  {journal} {\bibinfo  {journal}
  {Rept. Prog. Phys.}\ }\textbf {\bibinfo {volume} {79}},\ \bibinfo {pages}
  {046902} (\bibinfo {year} {2016})},\ \Eprint
  {http://arxiv.org/abs/1504.04623} {arXiv:1504.04623 [astro-ph.CO]}
  \BibitemShut {NoStop}%
\bibitem [{\citenamefont {Aghanim}\ \emph {et~al.}(2020)\citenamefont {Aghanim}
  \emph {et~al.}}]{Planck:2018vyg}%
  \BibitemOpen
  \bibfield  {author} {\bibinfo {author} {\bibfnamefont {N.}~\bibnamefont
  {Aghanim}} \emph {et~al.} (\bibinfo {collaboration} {Planck}),\ }\bibfield
  {title} {\enquote {\bibinfo {title} {{Planck 2018 results. VI. Cosmological
  parameters}},}\ }\href {\doibase 10.1051/0004-6361/201833910} {\bibfield
  {journal} {\bibinfo  {journal} {Astron. Astrophys.}\ }\textbf {\bibinfo
  {volume} {641}},\ \bibinfo {pages} {A6} (\bibinfo {year} {2020})},\ \Eprint
  {http://arxiv.org/abs/1807.06209} {arXiv:1807.06209 [astro-ph.CO]}
  \BibitemShut {NoStop}%
\bibitem [{\citenamefont {Riess}\ \emph
  {et~al.}(2021{\natexlab{a}})\citenamefont {Riess} \emph
  {et~al.}}]{Riess:2021jrx}%
  \BibitemOpen
  \bibfield  {author} {\bibinfo {author} {\bibfnamefont {Adam~G.}\ \bibnamefont
  {Riess}} \emph {et~al.},\ }\bibfield  {title} {\enquote {\bibinfo {title} {{A
  Comprehensive Measurement of the Local Value of the Hubble Constant with 1
  km/s/Mpc Uncertainty from the Hubble Space Telescope and the SH0ES Team}},}\
  }\href@noop {} {\  (\bibinfo {year} {2021}{\natexlab{a}})},\ \Eprint
  {http://arxiv.org/abs/2112.04510} {arXiv:2112.04510 [astro-ph.CO]}
  \BibitemShut {NoStop}%
\bibitem [{\citenamefont {Abdalla}\ \emph {et~al.}(2022)\citenamefont {Abdalla}
  \emph {et~al.}}]{Abdalla:2022yfr}%
  \BibitemOpen
  \bibfield  {author} {\bibinfo {author} {\bibfnamefont {Elcio}\ \bibnamefont
  {Abdalla}} \emph {et~al.},\ }\bibfield  {title} {\enquote {\bibinfo {title}
  {{Cosmology intertwined: A review of the particle physics, astrophysics, and
  cosmology associated with the cosmological tensions and anomalies}},}\ }\href
  {\doibase 10.1016/j.jheap.2022.04.002} {\bibfield  {journal} {\bibinfo
  {journal} {JHEAp}\ }\textbf {\bibinfo {volume} {34}},\ \bibinfo {pages}
  {49--211} (\bibinfo {year} {2022})},\ \Eprint
  {http://arxiv.org/abs/2203.06142} {arXiv:2203.06142 [astro-ph.CO]}
  \BibitemShut {NoStop}%
\bibitem [{\citenamefont {Freedman}\ \emph {et~al.}(2020)\citenamefont
  {Freedman}, \citenamefont {Madore}, \citenamefont {Hoyt}, \citenamefont
  {Jang}, \citenamefont {Beaton}, \citenamefont {Lee}, \citenamefont {Monson},
  \citenamefont {Neeley},\ and\ \citenamefont {Rich}}]{Freedman:2020dne}%
  \BibitemOpen
  \bibfield  {author} {\bibinfo {author} {\bibfnamefont {Wendy~L.}\
  \bibnamefont {Freedman}}, \bibinfo {author} {\bibfnamefont {Barry~F.}\
  \bibnamefont {Madore}}, \bibinfo {author} {\bibfnamefont {Taylor}\
  \bibnamefont {Hoyt}}, \bibinfo {author} {\bibfnamefont {In~Sung}\
  \bibnamefont {Jang}}, \bibinfo {author} {\bibfnamefont {Rachael}\
  \bibnamefont {Beaton}}, \bibinfo {author} {\bibfnamefont {Myung~Gyoon}\
  \bibnamefont {Lee}}, \bibinfo {author} {\bibfnamefont {Andrew}\ \bibnamefont
  {Monson}}, \bibinfo {author} {\bibfnamefont {Jill}\ \bibnamefont {Neeley}}, \
  and\ \bibinfo {author} {\bibfnamefont {Jeffrey}\ \bibnamefont {Rich}},\
  }\bibfield  {title} {\enquote {\bibinfo {title} {{Calibration of the Tip of
  the Red Giant Branch (TRGB)}},}\ }\href {\doibase 10.3847/1538-4357/ab7339}
  {\  (\bibinfo {year} {2020}),\ 10.3847/1538-4357/ab7339},\ \Eprint
  {http://arxiv.org/abs/2002.01550} {arXiv:2002.01550 [astro-ph.GA]}
  \BibitemShut {NoStop}%
\bibitem [{\citenamefont {Freedman}(2021)}]{Freedman:2021ahq}%
  \BibitemOpen
  \bibfield  {author} {\bibinfo {author} {\bibfnamefont {Wendy~L.}\
  \bibnamefont {Freedman}},\ }\bibfield  {title} {\enquote {\bibinfo {title}
  {{Measurements of the Hubble Constant: Tensions in Perspective}},}\ }\href
  {\doibase 10.3847/1538-4357/ac0e95} {\bibfield  {journal} {\bibinfo
  {journal} {Astrophys. J.}\ }\textbf {\bibinfo {volume} {919}},\ \bibinfo
  {pages} {16} (\bibinfo {year} {2021})},\ \Eprint
  {http://arxiv.org/abs/2106.15656} {arXiv:2106.15656 [astro-ph.CO]}
  \BibitemShut {NoStop}%
\bibitem [{\citenamefont {Abbott}\ \emph {et~al.}(2021)\citenamefont {Abbott}
  \emph {et~al.}}]{DES:2021wwk}%
  \BibitemOpen
  \bibfield  {author} {\bibinfo {author} {\bibfnamefont {T.~M.~C.}\
  \bibnamefont {Abbott}} \emph {et~al.} (\bibinfo {collaboration} {DES}),\
  }\bibfield  {title} {\enquote {\bibinfo {title} {{Dark Energy Survey Year 3
  Results: Cosmological Constraints from Galaxy Clustering and Weak
  Lensing}},}\ }\href@noop {} {\  (\bibinfo {year} {2021})},\ \Eprint
  {http://arxiv.org/abs/2105.13549} {arXiv:2105.13549 [astro-ph.CO]}
  \BibitemShut {NoStop}%
\bibitem [{\citenamefont {Asgari}\ \emph {et~al.}(2020)\citenamefont {Asgari}
  \emph {et~al.}}]{Asgari:2020wuj}%
  \BibitemOpen
  \bibfield  {author} {\bibinfo {author} {\bibfnamefont {Marika}\ \bibnamefont
  {Asgari}} \emph {et~al.} (\bibinfo {collaboration} {KiDS}),\ }\bibfield
  {title} {\enquote {\bibinfo {title} {{KiDS-1000 Cosmology: Cosmic shear
  constraints and comparison between two point statistics}},}\ \ }(\bibinfo
  {year} {2020})\ \Eprint {http://arxiv.org/abs/2007.15633} {arXiv:2007.15633
  [astro-ph.CO]} \BibitemShut {NoStop}%
\bibitem [{\citenamefont {Hikage}\ \emph {et~al.}(2019)\citenamefont {Hikage}
  \emph {et~al.}}]{HSC:2018mrq}%
  \BibitemOpen
  \bibfield  {author} {\bibinfo {author} {\bibfnamefont {Chiaki}\ \bibnamefont
  {Hikage}} \emph {et~al.} (\bibinfo {collaboration} {HSC}),\ }\bibfield
  {title} {\enquote {\bibinfo {title} {{Cosmology from cosmic shear power
  spectra with Subaru Hyper Suprime-Cam first-year data}},}\ }\href {\doibase
  10.1093/pasj/psz010} {\bibfield  {journal} {\bibinfo  {journal} {Publ.
  Astron. Soc. Jap.}\ }\textbf {\bibinfo {volume} {71}},\ \bibinfo {pages} {43}
  (\bibinfo {year} {2019})},\ \Eprint {http://arxiv.org/abs/1809.09148}
  {arXiv:1809.09148 [astro-ph.CO]} \BibitemShut {NoStop}%
\bibitem [{\citenamefont {Efstathiou}\ and\ \citenamefont
  {Lemos}(2018)}]{Efstathiou:2017rgv}%
  \BibitemOpen
  \bibfield  {author} {\bibinfo {author} {\bibfnamefont {George}\ \bibnamefont
  {Efstathiou}}\ and\ \bibinfo {author} {\bibfnamefont {Pablo}\ \bibnamefont
  {Lemos}},\ }\bibfield  {title} {\enquote {\bibinfo {title} {{Statistical
  inconsistencies in the KiDS-450 data set}},}\ }\href {\doibase
  10.1093/mnras/sty099} {\bibfield  {journal} {\bibinfo  {journal} {Mon. Not.
  Roy. Astron. Soc.}\ }\textbf {\bibinfo {volume} {476}},\ \bibinfo {pages}
  {151--157} (\bibinfo {year} {2018})},\ \Eprint
  {http://arxiv.org/abs/1707.00483} {arXiv:1707.00483 [astro-ph.CO]}
  \BibitemShut {NoStop}%
\bibitem [{\citenamefont {Efstathiou}(2020)}]{Efstathiou:2020wxn}%
  \BibitemOpen
  \bibfield  {author} {\bibinfo {author} {\bibfnamefont {G.}~\bibnamefont
  {Efstathiou}},\ }\bibfield  {title} {\enquote {\bibinfo {title} {{A Lockdown
  Perspective on the Hubble Tension (with comments from the SH0ES team)}},}\
  }\href@noop {} {\  (\bibinfo {year} {2020})},\ \Eprint
  {http://arxiv.org/abs/2007.10716} {arXiv:2007.10716 [astro-ph.CO]}
  \BibitemShut {NoStop}%
\bibitem [{\citenamefont {Will}(2014)}]{Will:2014kxa}%
  \BibitemOpen
  \bibfield  {author} {\bibinfo {author} {\bibfnamefont {Clifford~M.}\
  \bibnamefont {Will}},\ }\bibfield  {title} {\enquote {\bibinfo {title} {{The
  Confrontation between General Relativity and Experiment}},}\ }\href {\doibase
  10.12942/lrr-2014-4} {\bibfield  {journal} {\bibinfo  {journal} {Living Rev.
  Rel.}\ }\textbf {\bibinfo {volume} {17}},\ \bibinfo {pages} {4} (\bibinfo
  {year} {2014})},\ \Eprint {http://arxiv.org/abs/1403.7377} {arXiv:1403.7377
  [gr-qc]} \BibitemShut {NoStop}%
\bibitem [{\citenamefont {Abbott}\ \emph {et~al.}(2016)\citenamefont {Abbott}
  \emph {et~al.}}]{Abbott:2016blz}%
  \BibitemOpen
  \bibfield  {author} {\bibinfo {author} {\bibfnamefont {B.~P.}\ \bibnamefont
  {Abbott}} \emph {et~al.} (\bibinfo {collaboration} {Virgo, LIGO
  Scientific}),\ }\bibfield  {title} {\enquote {\bibinfo {title} {{Observation
  of Gravitational Waves from a Binary Black Hole Merger}},}\ }\href {\doibase
  10.1103/PhysRevLett.116.061102} {\bibfield  {journal} {\bibinfo  {journal}
  {Phys. Rev. Lett.}\ }\textbf {\bibinfo {volume} {116}},\ \bibinfo {pages}
  {061102} (\bibinfo {year} {2016})},\ \Eprint
  {http://arxiv.org/abs/1602.03837} {arXiv:1602.03837 [gr-qc]} \BibitemShut
  {NoStop}%
\bibitem [{\citenamefont {Abbott}\ \emph
  {et~al.}(2017{\natexlab{a}})\citenamefont {Abbott} \emph
  {et~al.}}]{TheLIGOScientific:2017qsa}%
  \BibitemOpen
  \bibfield  {author} {\bibinfo {author} {\bibfnamefont {B.?P.}\ \bibnamefont
  {Abbott}} \emph {et~al.} (\bibinfo {collaboration} {Virgo, LIGO
  Scientific}),\ }\bibfield  {title} {\enquote {\bibinfo {title} {{GW170817:
  Observation of Gravitational Waves from a Binary Neutron Star Inspiral}},}\
  }\href {\doibase 10.1103/PhysRevLett.119.161101} {\bibfield  {journal}
  {\bibinfo  {journal} {Phys. Rev. Lett.}\ }\textbf {\bibinfo {volume} {119}},\
  \bibinfo {pages} {161101} (\bibinfo {year} {2017}{\natexlab{a}})},\ \Eprint
  {http://arxiv.org/abs/1710.05832} {arXiv:1710.05832 [gr-qc]} \BibitemShut
  {NoStop}%
\bibitem [{\citenamefont {Akiyama}\ \emph {et~al.}(2019)\citenamefont {Akiyama}
  \emph {et~al.}}]{EventHorizonTelescope:2019dse}%
  \BibitemOpen
  \bibfield  {author} {\bibinfo {author} {\bibfnamefont {Kazunori}\
  \bibnamefont {Akiyama}} \emph {et~al.} (\bibinfo {collaboration} {Event
  Horizon Telescope}),\ }\bibfield  {title} {\enquote {\bibinfo {title} {{First
  M87 Event Horizon Telescope Results. I. The Shadow of the Supermassive Black
  Hole}},}\ }\href {\doibase 10.3847/2041-8213/ab0ec7} {\bibfield  {journal}
  {\bibinfo  {journal} {Astrophys. J. Lett.}\ }\textbf {\bibinfo {volume}
  {875}},\ \bibinfo {pages} {L1} (\bibinfo {year} {2019})},\ \Eprint
  {http://arxiv.org/abs/1906.11238} {arXiv:1906.11238 [astro-ph.GA]}
  \BibitemShut {NoStop}%
\bibitem [{\citenamefont {Riess}\ \emph {et~al.}(1998)\citenamefont {Riess}
  \emph {et~al.}}]{Riess:1998cb}%
  \BibitemOpen
  \bibfield  {author} {\bibinfo {author} {\bibfnamefont {Adam~G.}\ \bibnamefont
  {Riess}} \emph {et~al.} (\bibinfo {collaboration} {Supernova Search Team}),\
  }\bibfield  {title} {\enquote {\bibinfo {title} {{Observational evidence from
  supernovae for an accelerating universe and a cosmological constant}},}\
  }\href {\doibase 10.1086/300499} {\bibfield  {journal} {\bibinfo  {journal}
  {Astron. J.}\ }\textbf {\bibinfo {volume} {116}},\ \bibinfo {pages}
  {1009--1038} (\bibinfo {year} {1998})},\ \Eprint
  {http://arxiv.org/abs/astro-ph/9805201} {arXiv:astro-ph/9805201 [astro-ph]}
  \BibitemShut {NoStop}%
\bibitem [{\citenamefont {Perlmutter}\ \emph {et~al.}(1999)\citenamefont
  {Perlmutter} \emph {et~al.}}]{Perlmutter:1998np}%
  \BibitemOpen
  \bibfield  {author} {\bibinfo {author} {\bibfnamefont {S.}~\bibnamefont
  {Perlmutter}} \emph {et~al.} (\bibinfo {collaboration} {Supernova Cosmology
  Project}),\ }\bibfield  {title} {\enquote {\bibinfo {title} {{Measurements of
  Omega and Lambda from 42 high redshift supernovae}},}\ }\href {\doibase
  10.1086/307221} {\bibfield  {journal} {\bibinfo  {journal} {Astrophys. J.}\
  }\textbf {\bibinfo {volume} {517}},\ \bibinfo {pages} {565--586} (\bibinfo
  {year} {1999})},\ \Eprint {http://arxiv.org/abs/astro-ph/9812133}
  {arXiv:astro-ph/9812133 [astro-ph]} \BibitemShut {NoStop}%
\bibitem [{\citenamefont {Burgess}(2015)}]{Burgess:2013ara}%
  \BibitemOpen
  \bibfield  {author} {\bibinfo {author} {\bibfnamefont {C.~P.}\ \bibnamefont
  {Burgess}},\ }\bibfield  {title} {\enquote {\bibinfo {title} {{The
  Cosmological Constant Problem: Why it's hard to get Dark Energy from
  Micro-physics}},}\ }in\ \href {\doibase
  10.1093/acprof:oso/9780198728856.003.0004} {\emph {\bibinfo {booktitle}
  {{100e Ecole d'Ete de Physique: Post-Planck Cosmology Les Houches, France,
  July 8-August 2, 2013}}}}\ (\bibinfo {year} {2015})\ pp.\ \bibinfo {pages}
  {149--197},\ \Eprint {http://arxiv.org/abs/1309.4133} {arXiv:1309.4133
  [hep-th]} \BibitemShut {NoStop}%
\bibitem [{\citenamefont {Horndeski}(1974)}]{Horndeski:1974wa}%
  \BibitemOpen
  \bibfield  {author} {\bibinfo {author} {\bibfnamefont {Gregory~Walter}\
  \bibnamefont {Horndeski}},\ }\bibfield  {title} {\enquote {\bibinfo {title}
  {{Second-order scalar-tensor field equations in a four-dimensional space}},}\
  }\href {\doibase 10.1007/BF01807638} {\bibfield  {journal} {\bibinfo
  {journal} {Int. J. Theor. Phys.}\ }\textbf {\bibinfo {volume} {10}},\
  \bibinfo {pages} {363--384} (\bibinfo {year} {1974})}\BibitemShut {NoStop}%
\bibitem [{\citenamefont {Vainshtein}(1972)}]{Vainshtein:1972sx}%
  \BibitemOpen
  \bibfield  {author} {\bibinfo {author} {\bibfnamefont {A.~I.}\ \bibnamefont
  {Vainshtein}},\ }\bibfield  {title} {\enquote {\bibinfo {title} {{To the
  problem of nonvanishing gravitation mass}},}\ }\href {\doibase
  10.1016/0370-2693(72)90147-5} {\bibfield  {journal} {\bibinfo  {journal}
  {Phys. Lett.}\ }\textbf {\bibinfo {volume} {B39}},\ \bibinfo {pages}
  {393--394} (\bibinfo {year} {1972})}\BibitemShut {NoStop}%
\bibitem [{\citenamefont {Damour}\ and\ \citenamefont
  {Polyakov}(1994)}]{Damour:1994zq}%
  \BibitemOpen
  \bibfield  {author} {\bibinfo {author} {\bibfnamefont {T.}~\bibnamefont
  {Damour}}\ and\ \bibinfo {author} {\bibfnamefont {Alexander~M.}\ \bibnamefont
  {Polyakov}},\ }\bibfield  {title} {\enquote {\bibinfo {title} {{The String
  dilaton and a least coupling principle}},}\ }\href {\doibase
  10.1016/0550-3213(94)90143-0} {\bibfield  {journal} {\bibinfo  {journal}
  {Nucl. Phys.}\ }\textbf {\bibinfo {volume} {B423}},\ \bibinfo {pages}
  {532--558} (\bibinfo {year} {1994})},\ \Eprint
  {http://arxiv.org/abs/hep-th/9401069} {arXiv:hep-th/9401069 [hep-th]}
  \BibitemShut {NoStop}%
\bibitem [{\citenamefont {Khoury}\ and\ \citenamefont
  {Weltman}(2004)}]{Khoury:2003aq}%
  \BibitemOpen
  \bibfield  {author} {\bibinfo {author} {\bibfnamefont {Justin}\ \bibnamefont
  {Khoury}}\ and\ \bibinfo {author} {\bibfnamefont {Amanda}\ \bibnamefont
  {Weltman}},\ }\bibfield  {title} {\enquote {\bibinfo {title} {{Chameleon
  fields: Awaiting surprises for tests of gravity in space}},}\ }\href
  {\doibase 10.1103/PhysRevLett.93.171104} {\bibfield  {journal} {\bibinfo
  {journal} {Phys. Rev. Lett.}\ }\textbf {\bibinfo {volume} {93}},\ \bibinfo
  {pages} {171104} (\bibinfo {year} {2004})},\ \Eprint
  {http://arxiv.org/abs/astro-ph/0309300} {arXiv:astro-ph/0309300 [astro-ph]}
  \BibitemShut {NoStop}%
\bibitem [{\citenamefont {Hinterbichler}\ and\ \citenamefont
  {Khoury}(2010)}]{Hinterbichler:2010es}%
  \BibitemOpen
  \bibfield  {author} {\bibinfo {author} {\bibfnamefont {Kurt}\ \bibnamefont
  {Hinterbichler}}\ and\ \bibinfo {author} {\bibfnamefont {Justin}\
  \bibnamefont {Khoury}},\ }\bibfield  {title} {\enquote {\bibinfo {title}
  {{Symmetron Fields: Screening Long-Range Forces Through Local Symmetry
  Restoration}},}\ }\href {\doibase 10.1103/PhysRevLett.104.231301} {\bibfield
  {journal} {\bibinfo  {journal} {Phys. Rev. Lett.}\ }\textbf {\bibinfo
  {volume} {104}},\ \bibinfo {pages} {231301} (\bibinfo {year} {2010})},\
  \Eprint {http://arxiv.org/abs/1001.4525} {arXiv:1001.4525 [hep-th]}
  \BibitemShut {NoStop}%
\bibitem [{\citenamefont {Amendola}\ \emph {et~al.}(2008)\citenamefont
  {Amendola}, \citenamefont {Kunz},\ and\ \citenamefont
  {Sapone}}]{Amendola:2007rr}%
  \BibitemOpen
  \bibfield  {author} {\bibinfo {author} {\bibfnamefont {Luca}\ \bibnamefont
  {Amendola}}, \bibinfo {author} {\bibfnamefont {Martin}\ \bibnamefont {Kunz}},
  \ and\ \bibinfo {author} {\bibfnamefont {Domenico}\ \bibnamefont {Sapone}},\
  }\bibfield  {title} {\enquote {\bibinfo {title} {{Measuring the dark side
  (with weak lensing)}},}\ }\href {\doibase 10.1088/1475-7516/2008/04/013}
  {\bibfield  {journal} {\bibinfo  {journal} {JCAP}\ }\textbf {\bibinfo
  {volume} {0804}},\ \bibinfo {pages} {013} (\bibinfo {year} {2008})},\ \Eprint
  {http://arxiv.org/abs/0704.2421} {arXiv:0704.2421 [astro-ph]} \BibitemShut
  {NoStop}%
\bibitem [{\citenamefont {Bertschinger}\ and\ \citenamefont
  {Zukin}(2008)}]{Bertschinger:2008zb}%
  \BibitemOpen
  \bibfield  {author} {\bibinfo {author} {\bibfnamefont {Edmund}\ \bibnamefont
  {Bertschinger}}\ and\ \bibinfo {author} {\bibfnamefont {Phillip}\
  \bibnamefont {Zukin}},\ }\bibfield  {title} {\enquote {\bibinfo {title}
  {{Distinguishing Modified Gravity from Dark Energy}},}\ }\href {\doibase
  10.1103/PhysRevD.78.024015} {\bibfield  {journal} {\bibinfo  {journal} {Phys.
  Rev.}\ }\textbf {\bibinfo {volume} {D78}},\ \bibinfo {pages} {024015}
  (\bibinfo {year} {2008})},\ \Eprint {http://arxiv.org/abs/0801.2431}
  {arXiv:0801.2431 [astro-ph]} \BibitemShut {NoStop}%
\bibitem [{\citenamefont {Pogosian}\ \emph {et~al.}(2010)\citenamefont
  {Pogosian}, \citenamefont {Silvestri}, \citenamefont {Koyama},\ and\
  \citenamefont {Zhao}}]{Pogosian:2010tj}%
  \BibitemOpen
  \bibfield  {author} {\bibinfo {author} {\bibfnamefont {Levon}\ \bibnamefont
  {Pogosian}}, \bibinfo {author} {\bibfnamefont {Alessandra}\ \bibnamefont
  {Silvestri}}, \bibinfo {author} {\bibfnamefont {Kazuya}\ \bibnamefont
  {Koyama}}, \ and\ \bibinfo {author} {\bibfnamefont {Gong-Bo}\ \bibnamefont
  {Zhao}},\ }\bibfield  {title} {\enquote {\bibinfo {title} {{How to optimally
  parametrize deviations from General Relativity in the evolution of
  cosmological perturbations?}}}\ }\href {\doibase 10.1103/PhysRevD.81.104023}
  {\bibfield  {journal} {\bibinfo  {journal} {Phys. Rev.}\ }\textbf {\bibinfo
  {volume} {D81}},\ \bibinfo {pages} {104023} (\bibinfo {year} {2010})},\
  \Eprint {http://arxiv.org/abs/1002.2382} {arXiv:1002.2382 [astro-ph.CO]}
  \BibitemShut {NoStop}%
\bibitem [{\citenamefont {Gubitosi}\ \emph {et~al.}(2013)\citenamefont
  {Gubitosi}, \citenamefont {Piazza},\ and\ \citenamefont
  {Vernizzi}}]{Gubitosi:2012hu}%
  \BibitemOpen
  \bibfield  {author} {\bibinfo {author} {\bibfnamefont {Giulia}\ \bibnamefont
  {Gubitosi}}, \bibinfo {author} {\bibfnamefont {Federico}\ \bibnamefont
  {Piazza}}, \ and\ \bibinfo {author} {\bibfnamefont {Filippo}\ \bibnamefont
  {Vernizzi}},\ }\bibfield  {title} {\enquote {\bibinfo {title} {{The Effective
  Field Theory of Dark Energy}},}\ }\href {\doibase
  10.1088/1475-7516/2013/02/032} {\bibfield  {journal} {\bibinfo  {journal}
  {JCAP}\ }\textbf {\bibinfo {volume} {1302}},\ \bibinfo {pages} {032}
  (\bibinfo {year} {2013})},\ \bibinfo {note} {[JCAP1302,032(2013)]},\ \Eprint
  {http://arxiv.org/abs/1210.0201} {arXiv:1210.0201 [hep-th]} \BibitemShut
  {NoStop}%
\bibitem [{\citenamefont {Bloomfield}\ \emph {et~al.}(2013)\citenamefont
  {Bloomfield}, \citenamefont {Flanagan}, \citenamefont {Park},\ and\
  \citenamefont {Watson}}]{Bloomfield:2012ff}%
  \BibitemOpen
  \bibfield  {author} {\bibinfo {author} {\bibfnamefont {Jolyon~K.}\
  \bibnamefont {Bloomfield}}, \bibinfo {author} {\bibfnamefont {Eanna~E.}\
  \bibnamefont {Flanagan}}, \bibinfo {author} {\bibfnamefont {Minjoon}\
  \bibnamefont {Park}}, \ and\ \bibinfo {author} {\bibfnamefont {Scott}\
  \bibnamefont {Watson}},\ }\bibfield  {title} {\enquote {\bibinfo {title}
  {{Dark energy or modified gravity? An effective field theory approach}},}\
  }\href {\doibase 10.1088/1475-7516/2013/08/010} {\bibfield  {journal}
  {\bibinfo  {journal} {JCAP}\ }\textbf {\bibinfo {volume} {1308}},\ \bibinfo
  {pages} {010} (\bibinfo {year} {2013})},\ \Eprint
  {http://arxiv.org/abs/1211.7054} {arXiv:1211.7054 [astro-ph.CO]} \BibitemShut
  {NoStop}%
\bibitem [{\citenamefont {Gleyzes}\ \emph {et~al.}(2015)\citenamefont
  {Gleyzes}, \citenamefont {Langlois},\ and\ \citenamefont
  {Vernizzi}}]{Gleyzes:2014rba}%
  \BibitemOpen
  \bibfield  {author} {\bibinfo {author} {\bibfnamefont {Jerome}\ \bibnamefont
  {Gleyzes}}, \bibinfo {author} {\bibfnamefont {David}\ \bibnamefont
  {Langlois}}, \ and\ \bibinfo {author} {\bibfnamefont {Filippo}\ \bibnamefont
  {Vernizzi}},\ }\bibfield  {title} {\enquote {\bibinfo {title} {{A unifying
  description of dark energy}},}\ }\href {\doibase 10.1142/S021827181443010X}
  {\bibfield  {journal} {\bibinfo  {journal} {Int. J. Mod. Phys.}\ }\textbf
  {\bibinfo {volume} {D23}},\ \bibinfo {pages} {1443010} (\bibinfo {year}
  {2015})},\ \Eprint {http://arxiv.org/abs/1411.3712} {arXiv:1411.3712
  [hep-th]} \BibitemShut {NoStop}%
\bibitem [{\citenamefont {Bellini}\ and\ \citenamefont
  {Sawicki}(2014)}]{Bellini:2014fua}%
  \BibitemOpen
  \bibfield  {author} {\bibinfo {author} {\bibfnamefont {Emilio}\ \bibnamefont
  {Bellini}}\ and\ \bibinfo {author} {\bibfnamefont {Ignacy}\ \bibnamefont
  {Sawicki}},\ }\bibfield  {title} {\enquote {\bibinfo {title} {{Maximal
  freedom at minimum cost: linear large-scale structure in general
  modifications of gravity}},}\ }\href {\doibase 10.1088/1475-7516/2014/07/050}
  {\bibfield  {journal} {\bibinfo  {journal} {JCAP}\ }\textbf {\bibinfo
  {volume} {1407}},\ \bibinfo {pages} {050} (\bibinfo {year} {2014})},\ \Eprint
  {http://arxiv.org/abs/1404.3713} {arXiv:1404.3713 [astro-ph.CO]} \BibitemShut
  {NoStop}%
\bibitem [{\citenamefont {Zhao}\ \emph {et~al.}(2009)\citenamefont {Zhao},
  \citenamefont {Pogosian}, \citenamefont {Silvestri},\ and\ \citenamefont
  {Zylberberg}}]{Zhao:2008bn}%
  \BibitemOpen
  \bibfield  {author} {\bibinfo {author} {\bibfnamefont {Gong-Bo}\ \bibnamefont
  {Zhao}}, \bibinfo {author} {\bibfnamefont {Levon}\ \bibnamefont {Pogosian}},
  \bibinfo {author} {\bibfnamefont {Alessandra}\ \bibnamefont {Silvestri}}, \
  and\ \bibinfo {author} {\bibfnamefont {Joel}\ \bibnamefont {Zylberberg}},\
  }\bibfield  {title} {\enquote {\bibinfo {title} {{Searching for modified
  growth patterns with tomographic surveys}},}\ }\href {\doibase
  10.1103/PhysRevD.79.083513} {\bibfield  {journal} {\bibinfo  {journal} {Phys.
  Rev.}\ }\textbf {\bibinfo {volume} {D79}},\ \bibinfo {pages} {083513}
  (\bibinfo {year} {2009})},\ \Eprint {http://arxiv.org/abs/0809.3791}
  {arXiv:0809.3791 [astro-ph]} \BibitemShut {NoStop}%
\bibitem [{\citenamefont {Hojjati}\ \emph {et~al.}(2011)\citenamefont
  {Hojjati}, \citenamefont {Pogosian},\ and\ \citenamefont
  {Zhao}}]{Hojjati:2011ix}%
  \BibitemOpen
  \bibfield  {author} {\bibinfo {author} {\bibfnamefont {Alireza}\ \bibnamefont
  {Hojjati}}, \bibinfo {author} {\bibfnamefont {Levon}\ \bibnamefont
  {Pogosian}}, \ and\ \bibinfo {author} {\bibfnamefont {Gong-Bo}\ \bibnamefont
  {Zhao}},\ }\bibfield  {title} {\enquote {\bibinfo {title} {{Testing gravity
  with CAMB and CosmoMC}},}\ }\href {\doibase 10.1088/1475-7516/2011/08/005}
  {\bibfield  {journal} {\bibinfo  {journal} {JCAP}\ }\textbf {\bibinfo
  {volume} {1108}},\ \bibinfo {pages} {005} (\bibinfo {year} {2011})},\ \Eprint
  {http://arxiv.org/abs/1106.4543} {arXiv:1106.4543 [astro-ph.CO]} \BibitemShut
  {NoStop}%
\bibitem [{\citenamefont {Hu}\ \emph {et~al.}(2014)\citenamefont {Hu},
  \citenamefont {Raveri}, \citenamefont {Frusciante},\ and\ \citenamefont
  {Silvestri}}]{Hu:2013twa}%
  \BibitemOpen
  \bibfield  {author} {\bibinfo {author} {\bibfnamefont {Bin}\ \bibnamefont
  {Hu}}, \bibinfo {author} {\bibfnamefont {Marco}\ \bibnamefont {Raveri}},
  \bibinfo {author} {\bibfnamefont {Noemi}\ \bibnamefont {Frusciante}}, \ and\
  \bibinfo {author} {\bibfnamefont {Alessandra}\ \bibnamefont {Silvestri}},\
  }\bibfield  {title} {\enquote {\bibinfo {title} {{Effective Field Theory of
  Cosmic Acceleration: an implementation in CAMB}},}\ }\href {\doibase
  10.1103/PhysRevD.89.103530} {\bibfield  {journal} {\bibinfo  {journal} {Phys.
  Rev.}\ }\textbf {\bibinfo {volume} {D89}},\ \bibinfo {pages} {103530}
  (\bibinfo {year} {2014})},\ \Eprint {http://arxiv.org/abs/1312.5742}
  {arXiv:1312.5742 [astro-ph.CO]} \BibitemShut {NoStop}%
\bibitem [{\citenamefont {Zumalacarregui}\ \emph {et~al.}(2016)\citenamefont
  {Zumalacarregui}, \citenamefont {Bellini}, \citenamefont {Sawicki},\ and\
  \citenamefont {Lesgourgues}}]{Zumalacarregui:2016pph}%
  \BibitemOpen
  \bibfield  {author} {\bibinfo {author} {\bibfnamefont {Miguel}\ \bibnamefont
  {Zumalacarregui}}, \bibinfo {author} {\bibfnamefont {Emilio}\ \bibnamefont
  {Bellini}}, \bibinfo {author} {\bibfnamefont {Ignacy}\ \bibnamefont
  {Sawicki}}, \ and\ \bibinfo {author} {\bibfnamefont {Julien}\ \bibnamefont
  {Lesgourgues}},\ }\bibfield  {title} {\enquote {\bibinfo {title} {{hi\_class:
  Horndeski in the Cosmic Linear Anisotropy Solving System}},}\ }\href@noop {}
  {\  (\bibinfo {year} {2016})},\ \Eprint {http://arxiv.org/abs/1605.06102}
  {arXiv:1605.06102 [astro-ph.CO]} \BibitemShut {NoStop}%
\bibitem [{\citenamefont {Song}\ \emph {et~al.}(2011)\citenamefont {Song},
  \citenamefont {Zhao}, \citenamefont {Bacon}, \citenamefont {Koyama},
  \citenamefont {Nichol},\ and\ \citenamefont {Pogosian}}]{Song:2010fg}%
  \BibitemOpen
  \bibfield  {author} {\bibinfo {author} {\bibfnamefont {Yong-Seon}\
  \bibnamefont {Song}}, \bibinfo {author} {\bibfnamefont {Gong-Bo}\
  \bibnamefont {Zhao}}, \bibinfo {author} {\bibfnamefont {David}\ \bibnamefont
  {Bacon}}, \bibinfo {author} {\bibfnamefont {Kazuya}\ \bibnamefont {Koyama}},
  \bibinfo {author} {\bibfnamefont {Robert~C.}\ \bibnamefont {Nichol}}, \ and\
  \bibinfo {author} {\bibfnamefont {Levon}\ \bibnamefont {Pogosian}},\
  }\bibfield  {title} {\enquote {\bibinfo {title} {{Complementarity of Weak
  Lensing and Peculiar Velocity Measurements in Testing General Relativity}},}\
  }\href {\doibase 10.1103/PhysRevD.84.083523} {\bibfield  {journal} {\bibinfo
  {journal} {Phys. Rev.}\ }\textbf {\bibinfo {volume} {D84}},\ \bibinfo {pages}
  {083523} (\bibinfo {year} {2011})},\ \Eprint {http://arxiv.org/abs/1011.2106}
  {arXiv:1011.2106 [astro-ph.CO]} \BibitemShut {NoStop}%
\bibitem [{\citenamefont {Saltas}\ \emph {et~al.}(2014)\citenamefont {Saltas},
  \citenamefont {Sawicki}, \citenamefont {Amendola},\ and\ \citenamefont
  {Kunz}}]{Saltas:2014dha}%
  \BibitemOpen
  \bibfield  {author} {\bibinfo {author} {\bibfnamefont {Ippocratis~D.}\
  \bibnamefont {Saltas}}, \bibinfo {author} {\bibfnamefont {Ignacy}\
  \bibnamefont {Sawicki}}, \bibinfo {author} {\bibfnamefont {Luca}\
  \bibnamefont {Amendola}}, \ and\ \bibinfo {author} {\bibfnamefont {Martin}\
  \bibnamefont {Kunz}},\ }\bibfield  {title} {\enquote {\bibinfo {title}
  {{Anisotropic Stress as a Signature of Nonstandard Propagation of
  Gravitational Waves}},}\ }\href {\doibase 10.1103/PhysRevLett.113.191101}
  {\bibfield  {journal} {\bibinfo  {journal} {Phys. Rev. Lett.}\ }\textbf
  {\bibinfo {volume} {113}},\ \bibinfo {pages} {191101} (\bibinfo {year}
  {2014})},\ \Eprint {http://arxiv.org/abs/1406.7139} {arXiv:1406.7139
  [astro-ph.CO]} \BibitemShut {NoStop}%
\bibitem [{\citenamefont {Pogosian}\ and\ \citenamefont
  {Silvestri}(2016)}]{Pogosian:2016pwr}%
  \BibitemOpen
  \bibfield  {author} {\bibinfo {author} {\bibfnamefont {Levon}\ \bibnamefont
  {Pogosian}}\ and\ \bibinfo {author} {\bibfnamefont {Alessandra}\ \bibnamefont
  {Silvestri}},\ }\bibfield  {title} {\enquote {\bibinfo {title} {{What can
  Cosmology tell us about Gravity? Constraining Horndeski with Sigma and
  Mu}},}\ }\href {\doibase 10.1103/PhysRevD.94.104014} {\bibfield  {journal}
  {\bibinfo  {journal} {Phys. Rev.}\ }\textbf {\bibinfo {volume} {D94}},\
  \bibinfo {pages} {104014} (\bibinfo {year} {2016})},\ \Eprint
  {http://arxiv.org/abs/1606.05339} {arXiv:1606.05339 [astro-ph.CO]}
  \BibitemShut {NoStop}%
\bibitem [{\citenamefont {Silvestri}\ \emph {et~al.}(2013)\citenamefont
  {Silvestri}, \citenamefont {Pogosian},\ and\ \citenamefont
  {Buniy}}]{Silvestri:2013ne}%
  \BibitemOpen
  \bibfield  {author} {\bibinfo {author} {\bibfnamefont {Alessandra}\
  \bibnamefont {Silvestri}}, \bibinfo {author} {\bibfnamefont {Levon}\
  \bibnamefont {Pogosian}}, \ and\ \bibinfo {author} {\bibfnamefont {Roman~V.}\
  \bibnamefont {Buniy}},\ }\bibfield  {title} {\enquote {\bibinfo {title}
  {{Practical approach to cosmological perturbations in modified gravity}},}\
  }\href {\doibase 10.1103/PhysRevD.87.104015} {\bibfield  {journal} {\bibinfo
  {journal} {Phys. Rev.}\ }\textbf {\bibinfo {volume} {D87}},\ \bibinfo {pages}
  {104015} (\bibinfo {year} {2013})},\ \Eprint {http://arxiv.org/abs/1302.1193}
  {arXiv:1302.1193 [astro-ph.CO]} \BibitemShut {NoStop}%
\bibitem [{\citenamefont {Espejo}\ \emph {et~al.}(2019)\citenamefont {Espejo},
  \citenamefont {Peirone}, \citenamefont {Raveri}, \citenamefont {Koyama},
  \citenamefont {Pogosian},\ and\ \citenamefont {Silvestri}}]{Espejo:2018hxa}%
  \BibitemOpen
  \bibfield  {author} {\bibinfo {author} {\bibfnamefont {Juan}\ \bibnamefont
  {Espejo}}, \bibinfo {author} {\bibfnamefont {Simone}\ \bibnamefont
  {Peirone}}, \bibinfo {author} {\bibfnamefont {Marco}\ \bibnamefont {Raveri}},
  \bibinfo {author} {\bibfnamefont {Kazuya}\ \bibnamefont {Koyama}}, \bibinfo
  {author} {\bibfnamefont {Levon}\ \bibnamefont {Pogosian}}, \ and\ \bibinfo
  {author} {\bibfnamefont {Alessandra}\ \bibnamefont {Silvestri}},\ }\bibfield
  {title} {\enquote {\bibinfo {title} {{Phenomenology of Large Scale Structure
  in scalar-tensor theories: joint prior covariance of $w_{\textrm{DE}}$,
  $\Sigma$ and $\mu$ in Horndeski}},}\ }\href {\doibase
  10.1103/PhysRevD.99.023512} {\bibfield  {journal} {\bibinfo  {journal} {Phys.
  Rev. D}\ }\textbf {\bibinfo {volume} {99}},\ \bibinfo {pages} {023512}
  (\bibinfo {year} {2019})},\ \Eprint {http://arxiv.org/abs/1809.01121}
  {arXiv:1809.01121 [astro-ph.CO]} \BibitemShut {NoStop}%
\bibitem [{\citenamefont {Calabrese}\ \emph {et~al.}(2008)\citenamefont
  {Calabrese}, \citenamefont {Slosar}, \citenamefont {Melchiorri},
  \citenamefont {Smoot},\ and\ \citenamefont {Zahn}}]{Calabrese:2008rt}%
  \BibitemOpen
  \bibfield  {author} {\bibinfo {author} {\bibfnamefont {Erminia}\ \bibnamefont
  {Calabrese}}, \bibinfo {author} {\bibfnamefont {Anze}\ \bibnamefont
  {Slosar}}, \bibinfo {author} {\bibfnamefont {Alessandro}\ \bibnamefont
  {Melchiorri}}, \bibinfo {author} {\bibfnamefont {George~F.}\ \bibnamefont
  {Smoot}}, \ and\ \bibinfo {author} {\bibfnamefont {Oliver}\ \bibnamefont
  {Zahn}},\ }\bibfield  {title} {\enquote {\bibinfo {title} {{Cosmic Microwave
  Weak lensing data as a test for the dark universe}},}\ }\href {\doibase
  10.1103/PhysRevD.77.123531} {\bibfield  {journal} {\bibinfo  {journal} {Phys.
  Rev. D}\ }\textbf {\bibinfo {volume} {77}},\ \bibinfo {pages} {123531}
  (\bibinfo {year} {2008})},\ \Eprint {http://arxiv.org/abs/0803.2309}
  {arXiv:0803.2309 [astro-ph]} \BibitemShut {NoStop}%
\bibitem [{\citenamefont {Gleyzes}\ \emph {et~al.}(2016)\citenamefont
  {Gleyzes}, \citenamefont {Langlois}, \citenamefont {Mancarella},\ and\
  \citenamefont {Vernizzi}}]{Gleyzes:2015rua}%
  \BibitemOpen
  \bibfield  {author} {\bibinfo {author} {\bibfnamefont {Jerome}\ \bibnamefont
  {Gleyzes}}, \bibinfo {author} {\bibfnamefont {David}\ \bibnamefont
  {Langlois}}, \bibinfo {author} {\bibfnamefont {Michele}\ \bibnamefont
  {Mancarella}}, \ and\ \bibinfo {author} {\bibfnamefont {Filippo}\
  \bibnamefont {Vernizzi}},\ }\bibfield  {title} {\enquote {\bibinfo {title}
  {{Effective Theory of Dark Energy at Redshift Survey Scales}},}\ }\href
  {\doibase 10.1088/1475-7516/2016/02/056} {\bibfield  {journal} {\bibinfo
  {journal} {JCAP}\ }\textbf {\bibinfo {volume} {1602}},\ \bibinfo {pages}
  {056} (\bibinfo {year} {2016})},\ \Eprint {http://arxiv.org/abs/1509.02191}
  {arXiv:1509.02191 [astro-ph.CO]} \BibitemShut {NoStop}%
\bibitem [{\citenamefont {Abbott}\ \emph
  {et~al.}(2017{\natexlab{b}})\citenamefont {Abbott} \emph
  {et~al.}}]{Monitor:2017mdv}%
  \BibitemOpen
  \bibfield  {author} {\bibinfo {author} {\bibfnamefont {B.~P.}\ \bibnamefont
  {Abbott}} \emph {et~al.} (\bibinfo {collaboration} {Virgo, Fermi-GBM,
  INTEGRAL, LIGO Scientific}),\ }\bibfield  {title} {\enquote {\bibinfo {title}
  {{Gravitational Waves and Gamma-rays from a Binary Neutron Star Merger:
  GW170817 and GRB 170817A}},}\ }\href {\doibase 10.3847/2041-8213/aa920c}
  {\bibfield  {journal} {\bibinfo  {journal} {Astrophys. J.}\ }\textbf
  {\bibinfo {volume} {848}},\ \bibinfo {pages} {L13} (\bibinfo {year}
  {2017}{\natexlab{b}})},\ \Eprint {http://arxiv.org/abs/1710.05834}
  {arXiv:1710.05834 [astro-ph.HE]} \BibitemShut {NoStop}%
\bibitem [{\citenamefont {Deffayet}\ \emph {et~al.}(2009)\citenamefont
  {Deffayet}, \citenamefont {Esposito-Farese},\ and\ \citenamefont
  {Vikman}}]{Deffayet:2009wt}%
  \BibitemOpen
  \bibfield  {author} {\bibinfo {author} {\bibfnamefont {C.}~\bibnamefont
  {Deffayet}}, \bibinfo {author} {\bibfnamefont {Gilles}\ \bibnamefont
  {Esposito-Farese}}, \ and\ \bibinfo {author} {\bibfnamefont {A.}~\bibnamefont
  {Vikman}},\ }\bibfield  {title} {\enquote {\bibinfo {title} {{Covariant
  Galileon}},}\ }\href {\doibase 10.1103/PhysRevD.79.084003} {\bibfield
  {journal} {\bibinfo  {journal} {Phys. Rev. D}\ }\textbf {\bibinfo {volume}
  {79}},\ \bibinfo {pages} {084003} (\bibinfo {year} {2009})},\ \Eprint
  {http://arxiv.org/abs/0901.1314} {arXiv:0901.1314 [hep-th]} \BibitemShut
  {NoStop}%
\bibitem [{\citenamefont {Deffayet}\ \emph {et~al.}(2010)\citenamefont
  {Deffayet}, \citenamefont {Pujolas}, \citenamefont {Sawicki},\ and\
  \citenamefont {Vikman}}]{Deffayet:2010qz}%
  \BibitemOpen
  \bibfield  {author} {\bibinfo {author} {\bibfnamefont {Cedric}\ \bibnamefont
  {Deffayet}}, \bibinfo {author} {\bibfnamefont {Oriol}\ \bibnamefont
  {Pujolas}}, \bibinfo {author} {\bibfnamefont {Ignacy}\ \bibnamefont
  {Sawicki}}, \ and\ \bibinfo {author} {\bibfnamefont {Alexander}\ \bibnamefont
  {Vikman}},\ }\bibfield  {title} {\enquote {\bibinfo {title} {{Imperfect Dark
  Energy from Kinetic Gravity Braiding}},}\ }\href {\doibase
  10.1088/1475-7516/2010/10/026} {\bibfield  {journal} {\bibinfo  {journal}
  {JCAP}\ }\textbf {\bibinfo {volume} {1010}},\ \bibinfo {pages} {026}
  (\bibinfo {year} {2010})},\ \Eprint {http://arxiv.org/abs/1008.0048}
  {arXiv:1008.0048 [hep-th]} \BibitemShut {NoStop}%
\bibitem [{\citenamefont {Linder}(2018)}]{Linder:2018jil}%
  \BibitemOpen
  \bibfield  {author} {\bibinfo {author} {\bibfnamefont {Eric~V.}\ \bibnamefont
  {Linder}},\ }\bibfield  {title} {\enquote {\bibinfo {title} {{No Slip
  Gravity}},}\ }\href {\doibase 10.1088/1475-7516/2018/03/005} {\bibfield
  {journal} {\bibinfo  {journal} {JCAP}\ }\textbf {\bibinfo {volume} {03}},\
  \bibinfo {pages} {005} (\bibinfo {year} {2018})},\ \Eprint
  {http://arxiv.org/abs/1801.01503} {arXiv:1801.01503 [astro-ph.CO]}
  \BibitemShut {NoStop}%
\bibitem [{\citenamefont {Peirone}\ \emph {et~al.}(2018)\citenamefont
  {Peirone}, \citenamefont {Koyama}, \citenamefont {Pogosian}, \citenamefont
  {Raveri},\ and\ \citenamefont {Silvestri}}]{Peirone:2017ywi}%
  \BibitemOpen
  \bibfield  {author} {\bibinfo {author} {\bibfnamefont {Simone}\ \bibnamefont
  {Peirone}}, \bibinfo {author} {\bibfnamefont {Kazuya}\ \bibnamefont
  {Koyama}}, \bibinfo {author} {\bibfnamefont {Levon}\ \bibnamefont
  {Pogosian}}, \bibinfo {author} {\bibfnamefont {Marco}\ \bibnamefont
  {Raveri}}, \ and\ \bibinfo {author} {\bibfnamefont {Alessandra}\ \bibnamefont
  {Silvestri}},\ }\bibfield  {title} {\enquote {\bibinfo {title} {{Large-scale
  structure phenomenology of viable Horndeski theories}},}\ }\href {\doibase
  10.1103/PhysRevD.97.043519} {\bibfield  {journal} {\bibinfo  {journal} {Phys.
  Rev.}\ }\textbf {\bibinfo {volume} {D97}},\ \bibinfo {pages} {043519}
  (\bibinfo {year} {2018})},\ \Eprint {http://arxiv.org/abs/1712.00444}
  {arXiv:1712.00444 [astro-ph.CO]} \BibitemShut {NoStop}%
\bibitem [{\citenamefont {Zucca}\ \emph {et~al.}(2019)\citenamefont {Zucca},
  \citenamefont {Pogosian}, \citenamefont {Silvestri},\ and\ \citenamefont
  {Zhao}}]{Zucca:2019xhg}%
  \BibitemOpen
  \bibfield  {author} {\bibinfo {author} {\bibfnamefont {Alex}\ \bibnamefont
  {Zucca}}, \bibinfo {author} {\bibfnamefont {Levon}\ \bibnamefont {Pogosian}},
  \bibinfo {author} {\bibfnamefont {Alessandra}\ \bibnamefont {Silvestri}}, \
  and\ \bibinfo {author} {\bibfnamefont {Gong-Bo}\ \bibnamefont {Zhao}},\
  }\bibfield  {title} {\enquote {\bibinfo {title} {{MGCAMB with massive
  neutrinos and dynamical dark energy}},}\ }\href {\doibase
  10.1088/1475-7516/2019/05/001} {\bibfield  {journal} {\bibinfo  {journal}
  {JCAP}\ }\textbf {\bibinfo {volume} {05}},\ \bibinfo {pages} {001} (\bibinfo
  {year} {2019})},\ \Eprint {http://arxiv.org/abs/1901.05956} {arXiv:1901.05956
  [astro-ph.CO]} \BibitemShut {NoStop}%
\bibitem [{\citenamefont {Lewis}\ and\ \citenamefont
  {Bridle}(2002)}]{Lewis:2002ah}%
  \BibitemOpen
  \bibfield  {author} {\bibinfo {author} {\bibfnamefont {Antony}\ \bibnamefont
  {Lewis}}\ and\ \bibinfo {author} {\bibfnamefont {Sarah}\ \bibnamefont
  {Bridle}},\ }\bibfield  {title} {\enquote {\bibinfo {title} {{Cosmological
  parameters from CMB and other data: a Monte- Carlo approach}},}\ }\href@noop
  {} {\bibfield  {journal} {\bibinfo  {journal} {Phys. Rev.}\ }\textbf
  {\bibinfo {volume} {D66}},\ \bibinfo {pages} {103511} (\bibinfo {year}
  {2002})},\ \Eprint {http://arxiv.org/abs/astro-ph/0205436} {astro-ph/0205436}
  \BibitemShut {NoStop}%
\bibitem [{\citenamefont {Aghanim}\ \emph {et~al.}(2019)\citenamefont {Aghanim}
  \emph {et~al.}}]{Aghanim:2019ame}%
  \BibitemOpen
  \bibfield  {author} {\bibinfo {author} {\bibfnamefont {N.}~\bibnamefont
  {Aghanim}} \emph {et~al.} (\bibinfo {collaboration} {Planck}),\ }\bibfield
  {title} {\enquote {\bibinfo {title} {{Planck 2018 results. V. CMB power
  spectra and likelihoods}},}\ }\href@noop {} {\  (\bibinfo {year} {2019})},\
  \Eprint {http://arxiv.org/abs/1907.12875} {arXiv:1907.12875 [astro-ph.CO]}
  \BibitemShut {NoStop}%
\bibitem [{\citenamefont {Alam}\ \emph {et~al.}(2020)\citenamefont {Alam} \emph
  {et~al.}}]{Alam:2020sor}%
  \BibitemOpen
  \bibfield  {author} {\bibinfo {author} {\bibfnamefont {Shadab}\ \bibnamefont
  {Alam}} \emph {et~al.} (\bibinfo {collaboration} {eBOSS}),\ }\bibfield
  {title} {\enquote {\bibinfo {title} {{The Completed SDSS-IV extended Baryon
  Oscillation Spectroscopic Survey: Cosmological Implications from two Decades
  of Spectroscopic Surveys at the Apache Point observatory}},}\ }\href@noop {}
  {\  (\bibinfo {year} {2020})},\ \Eprint {http://arxiv.org/abs/2007.08991}
  {arXiv:2007.08991 [astro-ph.CO]} \BibitemShut {NoStop}%
\bibitem [{\citenamefont {Ross}\ \emph {et~al.}(2015)\citenamefont {Ross},
  \citenamefont {Samushia}, \citenamefont {Howlett}, \citenamefont {Percival},
  \citenamefont {Burden},\ and\ \citenamefont {Manera}}]{Ross:2014qpa}%
  \BibitemOpen
  \bibfield  {author} {\bibinfo {author} {\bibfnamefont {Ashley~J.}\
  \bibnamefont {Ross}}, \bibinfo {author} {\bibfnamefont {Lado}\ \bibnamefont
  {Samushia}}, \bibinfo {author} {\bibfnamefont {Cullan}\ \bibnamefont
  {Howlett}}, \bibinfo {author} {\bibfnamefont {Will~J.}\ \bibnamefont
  {Percival}}, \bibinfo {author} {\bibfnamefont {Angela}\ \bibnamefont
  {Burden}}, \ and\ \bibinfo {author} {\bibfnamefont {Marc}\ \bibnamefont
  {Manera}},\ }\bibfield  {title} {\enquote {\bibinfo {title} {{The clustering
  of the SDSS DR7 main Galaxy sample -- I. A 4 per cent distance measure at $z
  = 0.15$}},}\ }\href {\doibase 10.1093/mnras/stv154} {\bibfield  {journal}
  {\bibinfo  {journal} {Mon. Not. Roy. Astron. Soc.}\ }\textbf {\bibinfo
  {volume} {449}},\ \bibinfo {pages} {835--847} (\bibinfo {year} {2015})},\
  \Eprint {http://arxiv.org/abs/1409.3242} {arXiv:1409.3242 [astro-ph.CO]}
  \BibitemShut {NoStop}%
\bibitem [{\citenamefont {Beutler}\ \emph {et~al.}(2011)\citenamefont
  {Beutler}, \citenamefont {Blake}, \citenamefont {Colless}, \citenamefont
  {Jones}, \citenamefont {Staveley-Smith}, \citenamefont {Campbell},
  \citenamefont {Parker}, \citenamefont {Saunders},\ and\ \citenamefont
  {Watson}}]{Beutler:2011hx}%
  \BibitemOpen
  \bibfield  {author} {\bibinfo {author} {\bibfnamefont {Florian}\ \bibnamefont
  {Beutler}}, \bibinfo {author} {\bibfnamefont {Chris}\ \bibnamefont {Blake}},
  \bibinfo {author} {\bibfnamefont {Matthew}\ \bibnamefont {Colless}}, \bibinfo
  {author} {\bibfnamefont {D.Heath}\ \bibnamefont {Jones}}, \bibinfo {author}
  {\bibfnamefont {Lister}\ \bibnamefont {Staveley-Smith}}, \bibinfo {author}
  {\bibfnamefont {Lachlan}\ \bibnamefont {Campbell}}, \bibinfo {author}
  {\bibfnamefont {Quentin}\ \bibnamefont {Parker}}, \bibinfo {author}
  {\bibfnamefont {Will}\ \bibnamefont {Saunders}}, \ and\ \bibinfo {author}
  {\bibfnamefont {Fred}\ \bibnamefont {Watson}},\ }\bibfield  {title} {\enquote
  {\bibinfo {title} {{The 6dF Galaxy Survey: Baryon Acoustic Oscillations and
  the Local Hubble Constant}},}\ }\href {\doibase
  10.1111/j.1365-2966.2011.19250.x} {\bibfield  {journal} {\bibinfo  {journal}
  {Mon. Not. Roy. Astron. Soc.}\ }\textbf {\bibinfo {volume} {416}},\ \bibinfo
  {pages} {3017--3032} (\bibinfo {year} {2011})},\ \Eprint
  {http://arxiv.org/abs/1106.3366} {arXiv:1106.3366 [astro-ph.CO]} \BibitemShut
  {NoStop}%
\bibitem [{\citenamefont {Bautista}\ \emph {et~al.}(2020)\citenamefont
  {Bautista} \emph {et~al.}}]{Bautista:2020ahg}%
  \BibitemOpen
  \bibfield  {author} {\bibinfo {author} {\bibfnamefont {Julian~E.}\
  \bibnamefont {Bautista}} \emph {et~al.},\ }\bibfield  {title} {\enquote
  {\bibinfo {title} {{The Completed SDSS-IV extended Baryon Oscillation
  Spectroscopic Survey: measurement of the BAO and growth rate of structure of
  the luminous red galaxy sample from the anisotropic correlation function
  between redshifts 0.6 and 1}},}\ }\href {\doibase 10.1093/mnras/staa2800}
  {\bibfield  {journal} {\bibinfo  {journal} {Mon. Not. Roy. Astron. Soc.}\
  }\textbf {\bibinfo {volume} {500}},\ \bibinfo {pages} {736--762} (\bibinfo
  {year} {2020})},\ \Eprint {http://arxiv.org/abs/2007.08993} {arXiv:2007.08993
  [astro-ph.CO]} \BibitemShut {NoStop}%
\bibitem [{\citenamefont {de~Mattia}\ \emph {et~al.}(2021)\citenamefont
  {de~Mattia} \emph {et~al.}}]{deMattia:2020fkb}%
  \BibitemOpen
  \bibfield  {author} {\bibinfo {author} {\bibfnamefont {Arnaud}\ \bibnamefont
  {de~Mattia}} \emph {et~al.},\ }\bibfield  {title} {\enquote {\bibinfo {title}
  {{The Completed SDSS-IV extended Baryon Oscillation Spectroscopic Survey:
  measurement of the BAO and growth rate of structure of the emission line
  galaxy sample from the anisotropic power spectrum between redshift 0.6 and
  1.1}},}\ }\href {\doibase 10.1093/mnras/staa3891} {\bibfield  {journal}
  {\bibinfo  {journal} {Mon. Not. Roy. Astron. Soc.}\ }\textbf {\bibinfo
  {volume} {501}},\ \bibinfo {pages} {5616--5645} (\bibinfo {year} {2021})},\
  \Eprint {http://arxiv.org/abs/2007.09008} {arXiv:2007.09008 [astro-ph.CO]}
  \BibitemShut {NoStop}%
\bibitem [{\citenamefont {Hou}\ \emph {et~al.}(2020)\citenamefont {Hou} \emph
  {et~al.}}]{Hou:2020rse}%
  \BibitemOpen
  \bibfield  {author} {\bibinfo {author} {\bibfnamefont {Jiamin}\ \bibnamefont
  {Hou}} \emph {et~al.},\ }\bibfield  {title} {\enquote {\bibinfo {title} {{The
  Completed SDSS-IV extended Baryon Oscillation Spectroscopic Survey: BAO and
  RSD measurements from anisotropic clustering analysis of the Quasar Sample in
  configuration space between redshift 0.8 and 2.2}},}\ }\href {\doibase
  10.1093/mnras/staa3234} {\bibfield  {journal} {\bibinfo  {journal} {Mon. Not.
  Roy. Astron. Soc.}\ }\textbf {\bibinfo {volume} {500}},\ \bibinfo {pages}
  {1201--1221} (\bibinfo {year} {2020})},\ \Eprint
  {http://arxiv.org/abs/2007.08998} {arXiv:2007.08998 [astro-ph.CO]}
  \BibitemShut {NoStop}%
\bibitem [{\citenamefont {Neveux}\ \emph {et~al.}(2020)\citenamefont {Neveux}
  \emph {et~al.}}]{Neveux:2020voa}%
  \BibitemOpen
  \bibfield  {author} {\bibinfo {author} {\bibfnamefont {Richard}\ \bibnamefont
  {Neveux}} \emph {et~al.},\ }\bibfield  {title} {\enquote {\bibinfo {title}
  {{The completed SDSS-IV extended Baryon Oscillation Spectroscopic Survey: BAO
  and RSD measurements from the anisotropic power spectrum of the quasar sample
  between redshift 0.8 and 2.2}},}\ }\href {\doibase 10.1093/mnras/staa2780}
  {\bibfield  {journal} {\bibinfo  {journal} {Mon. Not. Roy. Astron. Soc.}\
  }\textbf {\bibinfo {volume} {499}},\ \bibinfo {pages} {210--229} (\bibinfo
  {year} {2020})},\ \Eprint {http://arxiv.org/abs/2007.08999} {arXiv:2007.08999
  [astro-ph.CO]} \BibitemShut {NoStop}%
\bibitem [{\citenamefont {Scolnic}\ \emph {et~al.}(2018)\citenamefont {Scolnic}
  \emph {et~al.}}]{Scolnic:2017caz}%
  \BibitemOpen
  \bibfield  {author} {\bibinfo {author} {\bibfnamefont {D.M.}\ \bibnamefont
  {Scolnic}} \emph {et~al.},\ }\bibfield  {title} {\enquote {\bibinfo {title}
  {{The Complete Light-curve Sample of Spectroscopically Confirmed SNe Ia from
  Pan-STARRS1 and Cosmological Constraints from the Combined Pantheon
  Sample}},}\ }\href {\doibase 10.3847/1538-4357/aab9bb} {\bibfield  {journal}
  {\bibinfo  {journal} {Astrophys. J.}\ }\textbf {\bibinfo {volume} {859}},\
  \bibinfo {pages} {101} (\bibinfo {year} {2018})},\ \Eprint
  {http://arxiv.org/abs/1710.00845} {arXiv:1710.00845 [astro-ph.CO]}
  \BibitemShut {NoStop}%
\bibitem [{\citenamefont {Abbott}\ \emph {et~al.}(2018)\citenamefont {Abbott}
  \emph {et~al.}}]{Abbott:2017wau}%
  \BibitemOpen
  \bibfield  {author} {\bibinfo {author} {\bibfnamefont {T.~M.~C.}\
  \bibnamefont {Abbott}} \emph {et~al.} (\bibinfo {collaboration} {DES}),\
  }\bibfield  {title} {\enquote {\bibinfo {title} {{Dark Energy Survey year 1
  results: Cosmological constraints from galaxy clustering and weak
  lensing}},}\ }\href {\doibase 10.1103/PhysRevD.98.043526} {\bibfield
  {journal} {\bibinfo  {journal} {Phys. Rev.}\ }\textbf {\bibinfo {volume}
  {D98}},\ \bibinfo {pages} {043526} (\bibinfo {year} {2018})},\ \Eprint
  {http://arxiv.org/abs/1708.01530} {arXiv:1708.01530 [astro-ph.CO]}
  \BibitemShut {NoStop}%
\bibitem [{\citenamefont {Riess}\ \emph
  {et~al.}(2021{\natexlab{b}})\citenamefont {Riess}, \citenamefont {Casertano},
  \citenamefont {Yuan}, \citenamefont {Bowers}, \citenamefont {Macri},
  \citenamefont {Zinn},\ and\ \citenamefont {Scolnic}}]{Riess:2020fzl}%
  \BibitemOpen
  \bibfield  {author} {\bibinfo {author} {\bibfnamefont {Adam~G.}\ \bibnamefont
  {Riess}}, \bibinfo {author} {\bibfnamefont {Stefano}\ \bibnamefont
  {Casertano}}, \bibinfo {author} {\bibfnamefont {Wenlong}\ \bibnamefont
  {Yuan}}, \bibinfo {author} {\bibfnamefont {J.~Bradley}\ \bibnamefont
  {Bowers}}, \bibinfo {author} {\bibfnamefont {Lucas}\ \bibnamefont {Macri}},
  \bibinfo {author} {\bibfnamefont {Joel~C.}\ \bibnamefont {Zinn}}, \ and\
  \bibinfo {author} {\bibfnamefont {Dan}\ \bibnamefont {Scolnic}},\ }\bibfield
  {title} {\enquote {\bibinfo {title} {{Cosmic Distances Calibrated to 1\%
  Precision with Gaia EDR3 Parallaxes and Hubble Space Telescope Photometry of
  75 Milky Way Cepheids Confirm Tension with $\Lambda$CDM}},}\ }\href {\doibase
  10.3847/2041-8213/abdbaf} {\bibfield  {journal} {\bibinfo  {journal}
  {Astrophys. J. Lett.}\ }\textbf {\bibinfo {volume} {908}},\ \bibinfo {pages}
  {6} (\bibinfo {year} {2021}{\natexlab{b}})},\ \Eprint
  {http://arxiv.org/abs/2012.08534} {arXiv:2012.08534 [astro-ph.CO]}
  \BibitemShut {NoStop}%
\bibitem [{\citenamefont {Lin}\ \emph {et~al.}(2019)\citenamefont {Lin},
  \citenamefont {Raveri},\ and\ \citenamefont {Hu}}]{Lin:2018nxe}%
  \BibitemOpen
  \bibfield  {author} {\bibinfo {author} {\bibfnamefont {Meng-Xiang}\
  \bibnamefont {Lin}}, \bibinfo {author} {\bibfnamefont {Marco}\ \bibnamefont
  {Raveri}}, \ and\ \bibinfo {author} {\bibfnamefont {Wayne}\ \bibnamefont
  {Hu}},\ }\bibfield  {title} {\enquote {\bibinfo {title} {{Phenomenology of
  Modified Gravity at Recombination}},}\ }\href {\doibase
  10.1103/PhysRevD.99.043514} {\bibfield  {journal} {\bibinfo  {journal} {Phys.
  Rev. D}\ }\textbf {\bibinfo {volume} {99}},\ \bibinfo {pages} {043514}
  (\bibinfo {year} {2019})},\ \Eprint {http://arxiv.org/abs/1810.02333}
  {arXiv:1810.02333 [astro-ph.CO]} \BibitemShut {NoStop}%
\bibitem [{\citenamefont {Crittenden}\ \emph {et~al.}(2012)\citenamefont
  {Crittenden}, \citenamefont {Zhao}, \citenamefont {Pogosian}, \citenamefont
  {Samushia},\ and\ \citenamefont {Zhang}}]{Crittenden:2011aa}%
  \BibitemOpen
  \bibfield  {author} {\bibinfo {author} {\bibfnamefont {Robert~G.}\
  \bibnamefont {Crittenden}}, \bibinfo {author} {\bibfnamefont {Gong-Bo}\
  \bibnamefont {Zhao}}, \bibinfo {author} {\bibfnamefont {Levon}\ \bibnamefont
  {Pogosian}}, \bibinfo {author} {\bibfnamefont {Lado}\ \bibnamefont
  {Samushia}}, \ and\ \bibinfo {author} {\bibfnamefont {Xinmin}\ \bibnamefont
  {Zhang}},\ }\bibfield  {title} {\enquote {\bibinfo {title} {{Fables of
  reconstruction: controlling bias in the dark energy equation of state}},}\
  }\href {\doibase 10.1088/1475-7516/2012/02/048} {\bibfield  {journal}
  {\bibinfo  {journal} {JCAP}\ }\textbf {\bibinfo {volume} {1202}},\ \bibinfo
  {pages} {048} (\bibinfo {year} {2012})},\ \Eprint
  {http://arxiv.org/abs/1112.1693} {arXiv:1112.1693 [astro-ph.CO]} \BibitemShut
  {NoStop}%
\bibitem [{\citenamefont {Crittenden}\ \emph {et~al.}(2009)\citenamefont
  {Crittenden}, \citenamefont {Pogosian},\ and\ \citenamefont
  {Zhao}}]{Crittenden:2005wj}%
  \BibitemOpen
  \bibfield  {author} {\bibinfo {author} {\bibfnamefont {Robert~G.}\
  \bibnamefont {Crittenden}}, \bibinfo {author} {\bibfnamefont {Levon}\
  \bibnamefont {Pogosian}}, \ and\ \bibinfo {author} {\bibfnamefont {Gong-Bo}\
  \bibnamefont {Zhao}},\ }\bibfield  {title} {\enquote {\bibinfo {title}
  {{Investigating dark energy experiments with principal components}},}\ }\href
  {\doibase 10.1088/1475-7516/2009/12/025} {\bibfield  {journal} {\bibinfo
  {journal} {JCAP}\ }\textbf {\bibinfo {volume} {0912}},\ \bibinfo {pages}
  {025} (\bibinfo {year} {2009})},\ \Eprint
  {http://arxiv.org/abs/astro-ph/0510293} {arXiv:astro-ph/0510293 [astro-ph]}
  \BibitemShut {NoStop}%
\bibitem [{\citenamefont {Wang}\ \emph {et~al.}(2012)\citenamefont {Wang},
  \citenamefont {Hui},\ and\ \citenamefont {Khoury}}]{Wang:2012kj}%
  \BibitemOpen
  \bibfield  {author} {\bibinfo {author} {\bibfnamefont {Junpu}\ \bibnamefont
  {Wang}}, \bibinfo {author} {\bibfnamefont {Lam}\ \bibnamefont {Hui}}, \ and\
  \bibinfo {author} {\bibfnamefont {Justin}\ \bibnamefont {Khoury}},\
  }\bibfield  {title} {\enquote {\bibinfo {title} {{No-Go Theorems for
  Generalized Chameleon Field Theories}},}\ }\href {\doibase
  10.1103/PhysRevLett.109.241301} {\bibfield  {journal} {\bibinfo  {journal}
  {Phys. Rev. Lett.}\ }\textbf {\bibinfo {volume} {109}},\ \bibinfo {pages}
  {241301} (\bibinfo {year} {2012})},\ \Eprint {http://arxiv.org/abs/1208.4612}
  {arXiv:1208.4612 [astro-ph.CO]} \BibitemShut {NoStop}%
\bibitem [{\citenamefont {Ade}\ \emph {et~al.}(2016)\citenamefont {Ade} \emph
  {et~al.}}]{Planck:2015bue}%
  \BibitemOpen
  \bibfield  {author} {\bibinfo {author} {\bibfnamefont {P.~A.~R.}\
  \bibnamefont {Ade}} \emph {et~al.} (\bibinfo {collaboration} {Planck}),\
  }\bibfield  {title} {\enquote {\bibinfo {title} {{Planck 2015 results. XIV.
  Dark energy and modified gravity}},}\ }\href {\doibase
  10.1051/0004-6361/201525814} {\bibfield  {journal} {\bibinfo  {journal}
  {Astron. Astrophys.}\ }\textbf {\bibinfo {volume} {594}},\ \bibinfo {pages}
  {A14} (\bibinfo {year} {2016})},\ \Eprint {http://arxiv.org/abs/1502.01590}
  {arXiv:1502.01590 [astro-ph.CO]} \BibitemShut {NoStop}%
\bibitem [{\citenamefont {Raveri}\ \emph {et~al.}()\citenamefont {Raveri},
  \citenamefont {Pogosian}, \citenamefont {Koyama}, \citenamefont {Martinelli},
  \citenamefont {Silvestri}, \citenamefont {Zhao}, \citenamefont {Li},\ and\
  \citenamefont {Peirone}}]{recon_prd}%
  \BibitemOpen
  \bibfield  {author} {\bibinfo {author} {\bibfnamefont {Marco}\ \bibnamefont
  {Raveri}}, \bibinfo {author} {\bibfnamefont {Levon}\ \bibnamefont
  {Pogosian}}, \bibinfo {author} {\bibfnamefont {Kazuya}\ \bibnamefont
  {Koyama}}, \bibinfo {author} {\bibfnamefont {Matteo}\ \bibnamefont
  {Martinelli}}, \bibinfo {author} {\bibfnamefont {Alessandra}\ \bibnamefont
  {Silvestri}}, \bibinfo {author} {\bibfnamefont {Gong-Bo}\ \bibnamefont
  {Zhao}}, \bibinfo {author} {\bibfnamefont {Jian}\ \bibnamefont {Li}}, \ and\
  \bibinfo {author} {\bibfnamefont {Simone}\ \bibnamefont {Peirone}},\
  }\bibfield  {title} {\enquote {\bibinfo {title} {{A joint reconstruction of
  the effective dark energy and modified growth evolution with and without a
  Horndeski prio}},}\ }\href@noop {} {\ }\Eprint
  {http://arxiv.org/abs/2107.12990} {arXiv:2107.12990 [astro-ph.CO]}
  \BibitemShut {NoStop}%
\bibitem [{\citenamefont {Pinho}\ \emph {et~al.}(2018)\citenamefont {Pinho},
  \citenamefont {Casas},\ and\ \citenamefont {Amendola}}]{Pinho:2018unz}%
  \BibitemOpen
  \bibfield  {author} {\bibinfo {author} {\bibfnamefont {Ana~Marta}\
  \bibnamefont {Pinho}}, \bibinfo {author} {\bibfnamefont {Santiago}\
  \bibnamefont {Casas}}, \ and\ \bibinfo {author} {\bibfnamefont {Luca}\
  \bibnamefont {Amendola}},\ }\bibfield  {title} {\enquote {\bibinfo {title}
  {{Model-independent reconstruction of the linear anisotropic stress
  $\eta$}},}\ }\href {\doibase 10.1088/1475-7516/2018/11/027} {\bibfield
  {journal} {\bibinfo  {journal} {JCAP}\ }\textbf {\bibinfo {volume} {11}},\
  \bibinfo {pages} {027} (\bibinfo {year} {2018})},\ \Eprint
  {http://arxiv.org/abs/1805.00027} {arXiv:1805.00027 [astro-ph.CO]}
  \BibitemShut {NoStop}%
\bibitem [{\citenamefont {Amendola}\ \emph {et~al.}(2013)\citenamefont
  {Amendola}, \citenamefont {Kunz}, \citenamefont {Motta}, \citenamefont
  {Saltas},\ and\ \citenamefont {Sawicki}}]{Amendola:2012ky}%
  \BibitemOpen
  \bibfield  {author} {\bibinfo {author} {\bibfnamefont {Luca}\ \bibnamefont
  {Amendola}}, \bibinfo {author} {\bibfnamefont {Martin}\ \bibnamefont {Kunz}},
  \bibinfo {author} {\bibfnamefont {Mariele}\ \bibnamefont {Motta}}, \bibinfo
  {author} {\bibfnamefont {Ippocratis~D.}\ \bibnamefont {Saltas}}, \ and\
  \bibinfo {author} {\bibfnamefont {Ignacy}\ \bibnamefont {Sawicki}},\
  }\bibfield  {title} {\enquote {\bibinfo {title} {{Observables and
  unobservables in dark energy cosmologies}},}\ }\href {\doibase
  10.1103/PhysRevD.87.023501} {\bibfield  {journal} {\bibinfo  {journal} {Phys.
  Rev. D}\ }\textbf {\bibinfo {volume} {87}},\ \bibinfo {pages} {023501}
  (\bibinfo {year} {2013})},\ \Eprint {http://arxiv.org/abs/1210.0439}
  {arXiv:1210.0439 [astro-ph.CO]} \BibitemShut {NoStop}%
\bibitem [{\citenamefont {Gleyzes}\ \emph {et~al.}(2013)\citenamefont
  {Gleyzes}, \citenamefont {Langlois}, \citenamefont {Piazza},\ and\
  \citenamefont {Vernizzi}}]{Gleyzes:2013ooa}%
  \BibitemOpen
  \bibfield  {author} {\bibinfo {author} {\bibfnamefont {Jerome}\ \bibnamefont
  {Gleyzes}}, \bibinfo {author} {\bibfnamefont {David}\ \bibnamefont
  {Langlois}}, \bibinfo {author} {\bibfnamefont {Federico}\ \bibnamefont
  {Piazza}}, \ and\ \bibinfo {author} {\bibfnamefont {Filippo}\ \bibnamefont
  {Vernizzi}},\ }\bibfield  {title} {\enquote {\bibinfo {title} {{Essential
  Building Blocks of Dark Energy}},}\ }\href {\doibase
  10.1088/1475-7516/2013/08/025} {\bibfield  {journal} {\bibinfo  {journal}
  {JCAP}\ }\textbf {\bibinfo {volume} {1308}},\ \bibinfo {pages} {025}
  (\bibinfo {year} {2013})},\ \Eprint {http://arxiv.org/abs/1304.4840}
  {arXiv:1304.4840 [hep-th]} \BibitemShut {NoStop}%
\bibitem [{\citenamefont {Raveri}(2020)}]{Raveri:2019mxg}%
  \BibitemOpen
  \bibfield  {author} {\bibinfo {author} {\bibfnamefont {Marco}\ \bibnamefont
  {Raveri}},\ }\bibfield  {title} {\enquote {\bibinfo {title} {{Reconstructing
  Gravity on Cosmological Scales}},}\ }\href {\doibase
  10.1103/PhysRevD.101.083524} {\bibfield  {journal} {\bibinfo  {journal}
  {Phys. Rev. D}\ }\textbf {\bibinfo {volume} {101}},\ \bibinfo {pages}
  {083524} (\bibinfo {year} {2020})},\ \Eprint
  {http://arxiv.org/abs/1902.01366} {arXiv:1902.01366 [astro-ph.CO]}
  \BibitemShut {NoStop}%
\bibitem [{\citenamefont {Park}\ \emph {et~al.}(2021)\citenamefont {Park},
  \citenamefont {Raveri},\ and\ \citenamefont {Jain}}]{Park:2021jmi}%
  \BibitemOpen
  \bibfield  {author} {\bibinfo {author} {\bibfnamefont {Minsu}\ \bibnamefont
  {Park}}, \bibinfo {author} {\bibfnamefont {Marco}\ \bibnamefont {Raveri}}, \
  and\ \bibinfo {author} {\bibfnamefont {Bhuvnesh}\ \bibnamefont {Jain}},\
  }\bibfield  {title} {\enquote {\bibinfo {title} {{Reconstructing
  Quintessence}},}\ }\href {\doibase 10.1103/PhysRevD.103.103530} {\bibfield
  {journal} {\bibinfo  {journal} {Phys. Rev. D}\ }\textbf {\bibinfo {volume}
  {103}},\ \bibinfo {pages} {103530} (\bibinfo {year} {2021})},\ \Eprint
  {http://arxiv.org/abs/2101.04666} {arXiv:2101.04666 [astro-ph.CO]}
  \BibitemShut {NoStop}%
\bibitem [{\citenamefont {Moss}\ \emph {et~al.}(2021)\citenamefont {Moss},
  \citenamefont {Copeland}, \citenamefont {Bamford},\ and\ \citenamefont
  {Clarke}}]{Moss:2021obd}%
  \BibitemOpen
  \bibfield  {author} {\bibinfo {author} {\bibfnamefont {Adam}\ \bibnamefont
  {Moss}}, \bibinfo {author} {\bibfnamefont {Edmund}\ \bibnamefont {Copeland}},
  \bibinfo {author} {\bibfnamefont {Steven}\ \bibnamefont {Bamford}}, \ and\
  \bibinfo {author} {\bibfnamefont {Thomas}\ \bibnamefont {Clarke}},\
  }\bibfield  {title} {\enquote {\bibinfo {title} {{A model-independent
  reconstruction of dark energy to very high redshift}},}\ }\href@noop {} {\
  (\bibinfo {year} {2021})},\ \Eprint {http://arxiv.org/abs/2109.14848}
  {arXiv:2109.14848 [astro-ph.CO]} \BibitemShut {NoStop}%
\bibitem [{\citenamefont {Zhao}\ \emph {et~al.}(2020)\citenamefont {Zhao} \emph
  {et~al.}}]{Zhao:2020tis}%
  \BibitemOpen
  \bibfield  {author} {\bibinfo {author} {\bibfnamefont {Gong-Bo}\ \bibnamefont
  {Zhao}} \emph {et~al.},\ }\bibfield  {title} {\enquote {\bibinfo {title}
  {{The Completed SDSS-IV extended Baryon Oscillation Spectroscopic Survey: a
  multi-tracer analysis in Fourier space for measuring the cosmic structure
  growth and expansion rate}},}\ }\href@noop {} {\  (\bibinfo {year} {2020})},\
  \Eprint {http://arxiv.org/abs/2007.09011} {arXiv:2007.09011 [astro-ph.CO]}
  \BibitemShut {NoStop}%
\bibitem [{\citenamefont {Wang}\ \emph {et~al.}(2020)\citenamefont {Wang} \emph
  {et~al.}}]{Wang:2020tje}%
  \BibitemOpen
  \bibfield  {author} {\bibinfo {author} {\bibfnamefont {Yuting}\ \bibnamefont
  {Wang}} \emph {et~al.},\ }\bibfield  {title} {\enquote {\bibinfo {title}
  {{The clustering of the SDSS-IV extended Baryon Oscillation Spectroscopic
  Survey DR16 luminous red galaxy and emission line galaxy samples: cosmic
  distance and structure growth measurements using multiple tracers in
  configuration space}},}\ }\href {\doibase 10.1093/mnras/staa2593} {\
  (\bibinfo {year} {2020}),\ 10.1093/mnras/staa2593},\ \Eprint
  {http://arxiv.org/abs/2007.09010} {arXiv:2007.09010 [astro-ph.CO]}
  \BibitemShut {NoStop}%
\bibitem [{\citenamefont {du~Mas~des Bourboux}\ \emph
  {et~al.}(2020)\citenamefont {du~Mas~des Bourboux} \emph
  {et~al.}}]{duMasdesBourboux:2020pck}%
  \BibitemOpen
  \bibfield  {author} {\bibinfo {author} {\bibfnamefont {Helion}\ \bibnamefont
  {du~Mas~des Bourboux}} \emph {et~al.},\ }\bibfield  {title} {\enquote
  {\bibinfo {title} {{The Completed SDSS-IV extended Baryon Oscillation
  Spectroscopic Survey: Baryon acoustic oscillations with Lyman-$\alpha$
  forests}},}\ }\href@noop {} {\  (\bibinfo {year} {2020})},\ \Eprint
  {http://arxiv.org/abs/2007.08995} {arXiv:2007.08995 [astro-ph.CO]}
  \BibitemShut {NoStop}%
\bibitem [{\citenamefont {Zhao}\ \emph {et~al.}(2017)\citenamefont {Zhao} \emph
  {et~al.}}]{BOSS:2016lpe}%
  \BibitemOpen
  \bibfield  {author} {\bibinfo {author} {\bibfnamefont {Gong-Bo}\ \bibnamefont
  {Zhao}} \emph {et~al.} (\bibinfo {collaboration} {BOSS}),\ }\bibfield
  {title} {\enquote {\bibinfo {title} {{The clustering of galaxies in the
  completed SDSS-III Baryon Oscillation Spectroscopic Survey: tomographic BAO
  analysis of DR12 combined sample in Fourier space}},}\ }\href {\doibase
  10.1093/mnras/stw3199} {\bibfield  {journal} {\bibinfo  {journal} {Mon. Not.
  Roy. Astron. Soc.}\ }\textbf {\bibinfo {volume} {466}},\ \bibinfo {pages}
  {762--779} (\bibinfo {year} {2017})},\ \Eprint
  {http://arxiv.org/abs/1607.03153} {arXiv:1607.03153 [astro-ph.CO]}
  \BibitemShut {NoStop}%
\bibitem [{\citenamefont {Benevento}\ \emph {et~al.}(2020)\citenamefont
  {Benevento}, \citenamefont {Hu},\ and\ \citenamefont
  {Raveri}}]{Benevento:2020fev}%
  \BibitemOpen
  \bibfield  {author} {\bibinfo {author} {\bibfnamefont {Giampaolo}\
  \bibnamefont {Benevento}}, \bibinfo {author} {\bibfnamefont {Wayne}\
  \bibnamefont {Hu}}, \ and\ \bibinfo {author} {\bibfnamefont {Marco}\
  \bibnamefont {Raveri}},\ }\bibfield  {title} {\enquote {\bibinfo {title}
  {{Can Late Dark Energy Transitions Raise the Hubble constant?}}}\ }\href
  {\doibase 10.1103/PhysRevD.101.103517} {\bibfield  {journal} {\bibinfo
  {journal} {Phys. Rev. D}\ }\textbf {\bibinfo {volume} {101}},\ \bibinfo
  {pages} {103517} (\bibinfo {year} {2020})},\ \Eprint
  {http://arxiv.org/abs/2002.11707} {arXiv:2002.11707 [astro-ph.CO]}
  \BibitemShut {NoStop}%
\bibitem [{\citenamefont {Barreira}\ \emph {et~al.}(2016)\citenamefont
  {Barreira}, \citenamefont {S\'anchez},\ and\ \citenamefont
  {Schmidt}}]{Barreira:2016ovx}%
  \BibitemOpen
  \bibfield  {author} {\bibinfo {author} {\bibfnamefont {Alexandre}\
  \bibnamefont {Barreira}}, \bibinfo {author} {\bibfnamefont {Ariel~G.}\
  \bibnamefont {S\'anchez}}, \ and\ \bibinfo {author} {\bibfnamefont {Fabian}\
  \bibnamefont {Schmidt}},\ }\bibfield  {title} {\enquote {\bibinfo {title}
  {{Validating estimates of the growth rate of structure with modified gravity
  simulations}},}\ }\href {\doibase 10.1103/PhysRevD.94.084022} {\bibfield
  {journal} {\bibinfo  {journal} {Phys. Rev. D}\ }\textbf {\bibinfo {volume}
  {94}},\ \bibinfo {pages} {084022} (\bibinfo {year} {2016})},\ \Eprint
  {http://arxiv.org/abs/1605.03965} {arXiv:1605.03965 [astro-ph.CO]}
  \BibitemShut {NoStop}%
\bibitem [{\citenamefont {Bose}\ \emph {et~al.}(2017)\citenamefont {Bose},
  \citenamefont {Koyama}, \citenamefont {Hellwing}, \citenamefont {Zhao},\ and\
  \citenamefont {Winther}}]{Bose:2017myh}%
  \BibitemOpen
  \bibfield  {author} {\bibinfo {author} {\bibfnamefont {Benjamin}\
  \bibnamefont {Bose}}, \bibinfo {author} {\bibfnamefont {Kazuya}\ \bibnamefont
  {Koyama}}, \bibinfo {author} {\bibfnamefont {Wojciech~A.}\ \bibnamefont
  {Hellwing}}, \bibinfo {author} {\bibfnamefont {Gong-Bo}\ \bibnamefont
  {Zhao}}, \ and\ \bibinfo {author} {\bibfnamefont {Hans~A.}\ \bibnamefont
  {Winther}},\ }\bibfield  {title} {\enquote {\bibinfo {title} {{Theoretical
  accuracy in cosmological growth estimation}},}\ }\href {\doibase
  10.1103/PhysRevD.96.023519} {\bibfield  {journal} {\bibinfo  {journal} {Phys.
  Rev. D}\ }\textbf {\bibinfo {volume} {96}},\ \bibinfo {pages} {023519}
  (\bibinfo {year} {2017})},\ \Eprint {http://arxiv.org/abs/1702.02348}
  {arXiv:1702.02348 [astro-ph.CO]} \BibitemShut {NoStop}%
\bibitem [{\citenamefont {Lewis}(2019)}]{Lewis:2019xzd}%
  \BibitemOpen
  \bibfield  {author} {\bibinfo {author} {\bibfnamefont {Antony}\ \bibnamefont
  {Lewis}},\ }\bibfield  {title} {\enquote {\bibinfo {title} {{GetDist: a
  Python package for analysing Monte Carlo samples}},}\ }\href
  {https://getdist.readthedocs.io} {\  (\bibinfo {year} {2019})},\ \Eprint
  {http://arxiv.org/abs/1910.13970} {arXiv:1910.13970 [astro-ph.IM]}
  \BibitemShut {NoStop}%
\end{thebibliography}%

\end{document}